\RequirePackage{fix-cm}

\documentclass[smallextended]{svjour3}       

\smartqed

\usepackage{changes}
\usepackage[english]{babel}
\usepackage{lipsum}
\usepackage{amsfonts}
\usepackage{dsfont}
\usepackage{graphicx}
\usepackage{epstopdf}
\usepackage{algorithmic}
\usepackage{xcolor}
\usepackage{array}
\usepackage{float}
\usepackage{mathrsfs}
\usepackage{mathtools}
\usepackage{mathdots}
\usepackage{multirow}
\usepackage{amstext}
\usepackage{amsmath}
\usepackage{amssymb}
\usepackage{enumitem}
\usepackage{amsopn}
\usepackage[T1]{fontenc}
\usepackage[latin9]{inputenc}
\usepackage{sansmathaccent}
\usepackage{lineno,hyperref}
\usepackage{multirow}
\usepackage{cite}
\usepackage[toc,page]{appendix}
\usepackage{arydshln}

\renewcommand{\PACSname}{\textbf{JEL classification}\enspace}

\definecolor{ao(english)}{rgb}{0.0, 0.5, 0.0}

\begin{document}

\title{Community structure in the World Trade Network based on communicability distances}

\titlerunning{Community structure in the WTN based on communicability distances}


\author{P. Bartesaghi \and G.P. Clemente \and \\ R. Grassi}

\institute{Paolo Bartesaghi \emph{Corresponding author.} \at
	Department of Statistics and Quantitative Methods, University of Milano - Bicocca, Via Bicocca degli Arcimboldi 8, 20126, Milano, Italy \\
	\email{paolo.bartesaghi@unimib.it} 
	\and
	Gian Paolo Clemente \at
	Department of Mathematics for Economics, Financial and Actuarial Sciences, Universit\`a Cattolica del Sacro Cuore, Largo Gemelli 1, 20123, Milano
	\email{gianpaolo.clemente@unicatt.it}
	\and
	Rosanna Grassi \at
	Department of Statistics and Quantitative Methods, University of Milano - Bicocca, Via Bicocca degli Arcimboldi 8, 20126, Milano, Italy \\
	\email{rosanna.grassi@unimib.it}}

\date{Received: date / Accepted: date}

\maketitle

\begin{abstract}
In this paper, we investigate the mesoscale structure of the World Trade Network. In this framework, a specific role is assumed by short and long-range interactions, and hence by the distance, between countries. %
Therefore, we identify clusters through a new procedure that exploits Estrada communicability distance and the vibrational communicability distance, 
which turn out to be particularly suitable for catching the inner structure of the economic network. The proposed methodology aims at finding the distance threshold that maximizes a specific modularity function defined for general metric spaces. Main advantages regard the computational efficiency of the procedure as well as the possibility to inspect intercluster and intracluster properties of the resulting communities.
The numerical analysis highlights peculiar relationships between countries and provides a rich set of information that can hardly be achieved within alternative clustering approaches.

\end{abstract}

\section{Introduction}
\label{sec:1}

International trade is based on a set of complex relationships between different countries. Both connections between countries and bilateral trade flows can be modelled as a dense network of interrelated and interconnected agents. A long-standing problem in this field is the detection of communities, namely subset of nodes among which the interactions are stronger than average. Indeed, the community structure of a network reveals how it is internally organized, highlighting the presence of special relationships between nodes, that might not be revealed by direct empirical analyses. 
\\
In this framework, a specific role is assumed by the distance between nodes. 
Indeed, the neighbours of a given node are immediately connected to such a node and they can affect its status 
most directly. Nonetheless, more distant nodes can influence this node while passing through intermediary ones. In the economic field, a network perspective is actually based on the idea that indirect trade relationships may be important (see, e.g., \cite{Fagiolo2015}). For instance, the authors in \cite{Abeysinghe2005} explain the impact of shocks on a given country by indirect trade links. Based on a global VaR approach, \cite{Dees2011} shows that countries that do not trade (very much) with the U.S. are largely influenced by its dominance over other trade partners linked with the U.S. via indirect spillovers. In \cite{Ward2013}, the bilateral trade is assumed not independent of the production, consumption, and trading decisions made by firms and consumers in third countries. A measure of the distance between nodes that also considers indirect connections is therefore crucial to catch deep interconnections between nodes. In this work, we will focus on two measures of distance or metrics on the network: the Estrada communicability distance \cite{Estrada2009} and the vibrational communicability distance \cite{Estrada2010book}. They both go beyond the limits of the immediate interaction between neighbours and they look simultaneously, albeit differently, at all the possible channels of interactions between nodes. The nearest two nodes are in each metric, the stronger is their interaction or, in other words, the higher is the level of communicability between them. \\

With this paper we contribute to the literature by proposing a specific methodology that exploits such metrics to inspect the mesoscale structure of the network, in search for strongly interacting clusters of nodes. Indeed, our purpose is twofold. We reveal hidden relationships between nodes due to non-immediate connections and long-range interactions and we show how this approach turns out to be particularly suitable when applied to a dense network like the World Trade Network (WTN).
More specifically, we exploits communicability and vibrational communicability metrics to group nodes whose mutuals distances are below a given threshold, i.e. whose interactions are stronger than a given value. Then we identify the optimal partition according to a maximum modularity criterion. It is well-known that modularity is a way to measure if a specific mesoscopic description of the network in terms of communities is more or less accurate. But, unlike the Girvan-Newman approach \cite{Newman2004}, we will refer to the modularity proposed in \cite{chang2016} for general metric spaces. In this way, we can exploit the additional information contained in the metric structure of the network. Among all the different partitions we get at different thresholds, we select the one providing the maximum modularity, according to the criterion described in \cite{chang2016}.
Our proposal is very efficient from a computational viewpoint. Indeed, given the specific distance matrix, the optimal solution can be easily evaluated varying the threshold. We cluster nodes going beyond the interactions between neighbours and considering all possible channels of interaction between them. We allow for a degree of flexibility by introducing a threshold. Varying the threshold, it is possible to depart from the optimal solution so that only the strongest (or the weakest) channels of communications emerge.

The paper is organised as follows. After a short review of the literature in Section \ref{sec:2}, main preliminaries and the definitions of the communicability functions are revised in Section \ref{sec:3}. These functions lead to two important metrics on networks, which are described in Section \ref{sec:5}. Section \ref{sec:6} contains the description of the proposed methodology, which is also tested on a suitable toy-model. In Section \ref{sec:7}, we apply our methodology to the World Trade Network. In particular, main characteristics of the network are described in Section \ref{sec:7:1}. The steps of the methodology are summarized in Section \ref{sec:7:2}. We report in Section \ref{sec:7:3} 
main results based on communicability and resistance distance, respectively. We show how the proposed methodology is able in capturing key economic clusters as well as in providing additional insights into intracluster and intercluster characteristics and of countries' relevance both in the community and in the whole network. Conclusions follow. Technical details are left in Appendices A and B.

\section{Literature Review}
\label{sec:2}

Community detection is an important topic in the analysis of the topological structure of complex systems. 
Its importance has grown over time in light of the remarkable progress in the description of large networks, together with the development of new powerful data analysis tools \cite{Fortunato2009}.  These advances have made it possible to extend the field of applicability of the theory not only to networks of enormous dimensions but also to weighted networks and direct networks \cite{Fagiolo2007,Rattigan2007, Clemente2018, Cerqueti2018}. Various methods and algorithms to detect communities on networks have been studied.  Some methods are algorithm-based, such as methods based on hierarchical clustering or edge removal \cite{Newman2004a}. Other methods are based on the optimization of specific criteria over all possible network partitions. In this context, it is well known the optimization of a modularity function according to Newman's definition \cite{Newman2004}. An exhaustive review about methods and algorithms can be found in \cite{Fortunato2010} and \cite{Fortunato2016}. Some authors proposed to detect communities by means of a quality measure called surprise \cite{Traag2015, Nicolini}. Inspired by this literature, recently  the authors in \cite{VaLidth} deal with detection of general mesoscale structures, such as core-periphery structures.

More recently the role of non-local interactions between nodes has been highlighted, that is interactions that do not exclusively involve the immediate neighbours of a given node. In particular, results connected to the idea of communicability introduced by Estrada in 2004 have proved to be extremely effective \cite{Estrada2005, Estrada2008, Estrada2009, Estrada2012}. All the more so by allowing a metric different from the shortest path metric to be introduced on the network. The purpose of this new metric is precisely to take into consideration long-range interactions between institutions. Some important similarities can be found between this new metric and the resistance distance, a well-known metric in network theory derived from the study of electric circuits \cite{Klein1993, Estrada2010book, Lee2019}, and its interpretation in terms of vibrational communicability \cite{Estrada2010paper, Bozzo2013, Ferraz2014, Mieghem2017}.

An area in which these concepts allow us to gain a deep insight into the hidden structures of the network is properly the WTN. The topology of the world trade web has been extensively analysed over time \cite{Serrano2003, Li2003, Garlaschelli2004, Garlaschelli2005, Garlaschelli2007, Fagiolo2008}. The behaviour of international trade flows, the impact of globalization on the international exchanges, the presence of a core-periphery structure or the evolution of the community centres of trade, are just some of the issues addressed by the recent developments \cite{Serrano2007, Tzekina2008, Fagiolo2010, DeBenedictis2011, Blochl2011}. Many works have dealt with the network from a multi layers perspective \cite{Snyder1979, Barigozzi2011} or aim to emphasize financial implications of the world trade or contagion processes on the network \cite{Wilhite2001, Reyes2008, Schiavo2010, Fagiolo2013, Fan2014, Varela2015, Giudici2016, DeBenedictis2016, Cepeda2019, Cerqueti2019}.

The impact of topology and metric properties on the stability and resilience of an economic or financial system has been widely studied in order to describe the large-scale pattern of dynamical processes inside the network \cite{Smith1992, Kali2007, Piccardi2018}. These processes determine the subsequent diversification of the export of a country, which can be compared with descriptive empirical indices of its potential growth, such as the one introduced in a very fruitful way in \cite{Hausmann2014}.

\section{Communicability in complex networks}
\label{sec:3}

The idea of communicability on a network is based on the ways in which a pair of nodes can communicate, namely through walks connecting them.
In the literature, two different definitions of communicability have been introduced: the Estrada Communicability and the Vibrational Communicability \cite{Estrada2008,Estrada2010paper}. We recall them in this section.

\subsection{Preliminary definitions}
\label{sec:3.1}
First of all, we briefly remind some preliminary  definitions. A network is formally represented by a graph ${\mathscr G}=(V,E)$ where $V$ and $E$ are the sets of $n$ nodes and $m$ edges, respectively.
Two nodes $i$ and $j$ are adjacent if there is an edge $(i,j)\in E$ connecting them.
The network is undirected if both $(i,j)$ and $(j,i)$ are elements of $E$. A $i-j$-path is a sequence of distinct vertices and edges between $i$ and $j$. 
The shortest path, or geodesic, between $i$ and $j$ is a path with the minimum number of edges. The length of a geodesic is called geodesic distance or shortest path distance $d(i,j)=d_{ij}$. A graph ${\mathscr G}$ is connected if, $\forall i,j \in V$, a $i-j$-path connecting them exists. \\

Adjacency relationships are represented by a binary symmetric matrix $\textbf{A}$ (adjacency matrix). Graphs considered here will be always connected and without loops; in this case $a_{ii}=0$ $\forall i=1,...,n$.
We denote with $\lambda_1\geq \lambda_2\geq \dots \geq \lambda_n$ the eigenvalues of $\textbf{A}$, and $\varphi_i,i=1,...,n$ the corresponding eigenvectors.\\ 
The degree $k_{i}$ of a node $i$ is the number of edges incident on it. 
The diagonal matrix whose diagonal entries are $k_i$ is  $\textbf{K}$. The Laplacian matrix is $\textbf{L}=\textbf{K}-\textbf{A}$. $\textbf{L}$ is a positive
semidefinite symmetric matrix. We denote the eigenvalues of  $\textbf{L}$ by $\mu_1\geq \mu_2\geq \dots > \mu_n=0$ and $\psi_i,i=1,...,n$ the corresponding eigenvectors. 

A graph ${\mathscr G}$ is weighted when a positive real number
$w_{ij}>0$ is associated with the edge $(i,j)$. 
We define the strength $s_i$ as the sum of the weights of the edges adjacent to $i$. 
The definition of geodesic path still holds, and it is a weighted path with the minimum sum of edge weights. 
In this case, the adjacency matrix is a non-negative symmetric matrix $\textbf{W}$. 
When $w_{ij}=1$ if $(i,j) \in E$, then the
graph is unweighted. Thus, the unweighted case can be viewed as a particular weighted one. 
 




\subsection{Estrada Communicability}
\label{sec:4.1}

The {\it Estrada communicability} \cite{Estrada2008} between two nodes $i$ and $j$ is defined as: 

\begin{equation}
\label{Ecomm}
	G_{ij}=\sum_{k=0}^{+\infty}\frac{1}{k!}[\textbf{A}^k]_{ij}=\left[ e^{\textbf{A}} \right]_{ij}.
\end{equation}

As the $ij$-entry of the $k$-power of the adjacency matrix $\textbf{A}$ counts the number of walks of length $k$ starting at $i$ and ending at $j$, $G_{ij}$ accounts for all channels of communication between two nodes, giving more weight to the shortest routes connecting them. It can also be interpreted as a measure of the probability that a particle starting at $i$ ends up at $j$ after wandering randomly on the complex network. The communicability matrix is denoted by $\textbf{G}$.

By definition,  it follows that $G_{ij} >0$. Moreover, $G_{ij}$ can be conveniently expressed using the spectral decomposition of $\textbf{A}$ as follows \cite{Estrada2008}:

$$
G_{ij }=\sum_{k=1}^{n}\varphi_{k}(i)\varphi_{k}(j)e^{\lambda_k},
$$

where $\varphi_{k}(i)$ is the $i$-component of the $k$-th eigenvector associated with $\lambda_k$.

It is worth noting that since $G_{ii}$ characterizes the importance of a node according to its participation in all closed walks starting and ending at it, we recover the so-called {\it subgraph centrality} (see \cite{Estrada2005}). 

In the case of a weighted network the communicability function is defined as

\begin{equation}
\label{wecomm}
G_{ij}=\sum_{k=0}^{+\infty}\frac{1}{k!}[(\textbf{S}^{-{1\over 2}}\textbf{W}\textbf{S}^{-{1\over 2}})^k]_{i j}=\left[ e^{(\textbf{S}^{-{1\over 2}}\textbf{W}\textbf{S}^{-{1\over 2}})} \right]_{ij}
\end{equation}

where $\textbf{S}$ is the diagonal matrix whose diagonal entries are the strengths of the nodes. We will call this quantity weighted communicability. The weighted communicability is particularly suitable to be applied to the study of input-output networks.



\subsection{Vibrational Communicability}
\label{sec:4.2}

Communicability can be alternatively defined through the following model from Physics.
Let us suppose that nodes of the network are objects of negligible identical mass connected by springs in a plane grid. Nodes can oscillate in the direction perpendicular to the plane and the displacement of the node $i$ from its rest position is $z_{i}$. The elastic force applied to node $i$ is given by $F_i ={\cal K}\sum_{j}A_{ij}(z_i-z_j)$, where $\cal K$ is the common elastic constant of each spring.
An elastic potential energy can be assigned to each perturbed spring and the potential energy of all the springs connected with node $i$ is given by $U_i=\frac{1}{2}\, {\cal K}\sum_{j}A_{ij}(z_i-z_j)^2$.\\
The overall potential energy of the network is therefore

\begin{equation}
\label{potential}
	U=\frac{1}{4}\, {\cal K}\sum_{i,j}A_{i j}(z_i-z_j)^2=\frac{1}{2}\, {\cal K}\sum_{i,j}z_{i}L_{ij} z_j
\end{equation}

where $L_{ij}$ is the $ij$-entry of $\textbf{L}$.

The reciprocal influence of two nodes $i$ and $j$ in their positions $z_i$ and $z_j$ is computed by means of the Green's function, according to the classical Boltzmann's distribution \cite{Estrada2010paper,Estrada2010book}. This mutual influence can be interpreted as the correlation function between the displacements $z$ of two nodes in the network:

$$
G^{v}_{ij}(\beta)=\left\langle z_i z_j\right\rangle =\frac{1}{\cal Z}\int z_i z_j e^{-\beta U} d\textbf{z}
$$

where $\beta$ is a constant and ${\cal Z}=\int e^{-\beta U} d\textbf{z}$ is the partition function. Using the non-zero eigenvalues of $\textbf{L}$, ${\cal Z}$ can be expressed as 

\begin{equation}
\label{partition}
\begin{split}
{\cal Z}&=\int e^{-\frac{1}{2}\beta {\cal K}\sum_{ij}z_{i}L_{ij} z_j}\prod_{k}dz_k=\prod_{k=1}^{n-1}\sqrt{\frac{2\pi}{\beta {\cal K}\mu_k}}
\end{split}
\end{equation}

so that the correlation function can be rewritten in the final form 

\begin{equation}
\label{vibrational2}
G^{v}_{ij}(\beta)=\sum_{k=1}^{n-1}\frac{\psi_{k}(i)\psi_{k}(j)}{\beta {\cal K}\mu_k}
\end{equation}

where $\psi_k$ is the eigenvector associated with $\mu_k$. Introducing the Moore-Penrose pseudo-inverse of the Laplacian  $\textbf{L}^+$ \cite{Bozzo2013,Gutman2004}, the vibrational communicability between nodes $i$ and $j$ is defined as

\begin{equation}
\label{vibrational}
G^{v}_{ij}(\beta)=\frac{1}{\beta {\cal K}}L^{+}_{ij}
\end{equation}


The vibrational communicability matrix is denoted by $\textbf{G}^v$. In the remainder of the paper we will assume $\beta=1$ and ${\cal K}=1$, so that $G^{v}_{ij}=L^{+}_{ij}$. 

\noindent The detailed computations for previous formulas are reported in Appendix A. \\

\section{Metrics on networks}
\label{sec:5}
Metric properties play an important role in the study of the structure and dynamics of networks. The best known metric is the so-called shortest path distance. 
In the literature other metrics have been defined, each one stressing different features of the network. We remind the definitions of communicability distance and resistance distance, in view of their following application to the WTN.

\subsection{Communicability Distance}
\label{sec:5.1}
The communicability distance $\xi_{ij}$ is defined as (see \cite{Estrada2012}):
\begin{equation}
	\label{communicabilitydistance}
	\xi_{i j}=G_{ii}-2G_{ij}+G_{jj}.
\end{equation}
As already observed, $G_{ii}$ is the subgraph centrality of $i$ and it measures the amount of information that starts from and returns to node $i$ after having wandered through the network. On the other hand, $G_{ij}$ measures the amount of information transmitted from $i$ to $j$.
Notice that the word {\it information} is meant in its broadest sense. Therefore, information flow can be any kind of flow along edges: money, current, traffic and so on. 
Thus, the quantity $\xi_{ij}$ accounts for the difference in the amount of information that returns to the nodes $i$ and $j$ and the amount of information exchanged between them. 

The greater is $G_{ij}$, the larger the information exchanged and the nearer are the nodes; the greater are $G_{ii}$ or $G_{jj}$, the larger the information that comes back to the nodes and the farther are the nodes. 
In a matrix form, $\xi_{i j}$ can be expressed as follows:

$$
{\bf \Xi}=\textbf{g}\textbf{u}^T-2\textbf{G}+\textbf{u}\textbf{g}^T
$$

where  $\textbf{g}=[G_{11},\dots, G_{nn}]^T$ is the vector of subgraph centralities and $\textbf{u}$ the all $1$'s $n-$vector. Since $\xi_{i j}$ is 
a metric, then $G_{ii}+G_{jj}\geq 2G_{ij}$, i.e., no matter what the structure of the network is, the amount of information absorbed by a pair of nodes is always larger than the amount of information transmitted between them. \\



\subsection{Resistance Distance}
\label{sec:5.2}
The vibrational communicability distance between $i$ and $j$ is defined as (see  \cite{Estrada2010book,Mieghem2017}): 

\begin{equation}
\label{resistancedistance}
\omega_{ij}=G^{v}_{ii}-2G^{v}_{ij}+G^{v}_{jj}.
\end{equation}

\noindent Formula \ref{resistancedistance} can be written in a more suitable way. Indeed, recalling that $G^{v}_{ij}=L^{+}_{ij}$, we have:

\begin{equation}
\begin{split}
\label{resistancedistance2}
\omega_{ij}
&=L^{+}_{ii}-2L^{+}_{ij}+L^{+}_{jj}\\
&=(\textbf{e}_i-\textbf{e}_j)^T{\textbf{L}^+}(\textbf{e}_i-\textbf{e}_j)\\
&=(\textbf{e}_i-\textbf{e}_j)^T\left[\left( \textbf{L}+ \frac{1}{n} \textbf{J} \right)^{-1} -\frac{1}{n}\textbf{J} \right](\textbf{e}_i-\textbf{e}_j)\\
&=
(\textbf{e}_i-\textbf{e}_j)^T\left( \textbf{L}+ \frac{1}{n} \textbf{J} \right)^{-1} (\textbf{e}_i-\textbf{e}_j)\\
\end{split}
\end{equation}

\noindent where $\textbf{e}_k$, $k=1, \dots, n$, is the standard basis in $\mathds R^n$ and $\textbf{J}=\textbf{u}\textbf{u}^{T}$ is the matrix whose entries are all $1$. Note that in the previous chain of equalities we made use of the following expression of the pseudo-inverse $\textbf{L}^+=\left( \textbf{L}+ \frac{1}{n} \textbf{J} \right)^{-1} -\frac{1}{n}\textbf{J}$, proved in \cite{Gutman2004}.

Equation \ref{resistancedistance2} offers an interesting interpretation of the resistance distance. We synthesize here the main idea, referring to Appendix B for a more detailed discussion.
Let ${\textbf v}=[v_1, v_2, \dots, v_n]^T$ be a vector representing attributes of the nodes -- for instance,  the Gross Domestic Product (GDP) of a country or the assets of a financial institution -- and suppose that there are currents or flows (of money, for instance) along the edges of the network.
The operator $\left( \textbf{L}+ \frac{1}{n} \textbf{J} \right)^{-1}$ allows to obtain the state vector that gives rise to a given set of flows. In formula \ref{resistancedistance2}, the vector $({\textbf e}_i-{\textbf e}_j)$ refers to a global flow equal to $+1$ from node $i$, a flow equal to $-1$ into node $j$ and a flow equal to $0$ for the other ones. When we apply  $\left( \textbf{L}+ \frac{1}{n} \textbf{J} \right)^{-1}$ to $({\textbf e}_i-{\textbf e}_j)$, we get the state vector ${\textbf v}=[v_1, v_2, \dots, v_n]^T$ of attributes on nodes that gives rise to these flows. Finally, the left inner product with $({\textbf e}_i-{\textbf e}_j)$ in formula \ref{resistancedistance2} gives $v_i-v_j$, namely, the difference between attributes of nodes $i$ and $j$. This gradient produces exactly the flow $+1$ from node $i$ and $-1$ to node $j$. 
If $v_i-v_j$ is big, we need a big difference in order to produce such a unit flow and so we have a big resistance between nodes $i$ and $j$. If $v_i-v_j$ is small, it is enough a low difference in order to produce such a unit flow and so we have a low resistance between nodes $i$ and $j$. If $\omega_{ij}$ is big we have a high resistance distance between $i$ and $j$. Therefore, these two nodes do not communicate easily. Vice versa a low value of $\omega_{ij}$ means a high level of communication between the nodes. $\omega_{ij}$ is called {\it effective resistance} between nodes $i$ and $j$ and ${\bf \Omega}=[\omega_{ij}]$ is the resistance matrix.

In literature, it is known an important close form for $\textbf{L}^+$ in terms of $\bf \Omega$:

$$
\textbf{L}^+=\frac{1}{2}\left[ \frac{1}{n}({\bf \Omega} \textbf{J}+\textbf{J}{\bf \Omega})-\frac{1}{n^2}\textbf{J}{\bf \Omega} \textbf{J}-{\bf \Omega} \right]
$$

which allows us to rewrite the diagonal elements of the matrix $\textbf{L}^+$ in a useful form\footnote{$L^{+}_{ii}=\frac{1}{2n}({\bf \Omega} \textbf{J})_{ii}+\frac{1}{2n}(\textbf{J}{\bf \Omega})_{ii}-\frac{1}{2n^2}(\textbf{J}{\bf \Omega} \textbf{J})_{ii}-\frac{1}{2}({\bf \Omega})_{ii}
=\frac{1}{2n}\sum_{j}\omega_{ij}J_{ji}+\frac{1}{2n}\sum_{j}J_{ij}\omega_{ji}-\frac{1}{2n^2}\sum_{jk}J_{ij}\omega_{jk}J_{ki}-0
=\frac{1}{2n}\sum_{j}\omega_{ij}+\frac{1}{2n}\sum_{j}\omega_{ij}-\frac{1}{2n^2}\sum_{jk}\omega_{jk}
=\frac{1}{n}\sum_{j}\omega_{ij}-\frac{R}{n^2}$}

\begin{equation*}
L^{+}_{ii}=\frac{1}{n}\sum_{j}\omega_{ij}-\frac{R}{n^2}
\end{equation*}

where 

$$
R=\frac{1}{2}\sum_{i,j}\omega_{ij}=\sum_{i=1}^{n}\sum_{j=i+1}^{n}\omega_{ij}=\frac{1}{2}\textbf{u}^T {\bf \Omega}\textbf{u} = n\, {\rm tr}\, \textbf{L}^{+}=n \sum_{k=1}^{n-1}\frac{1}{\mu_k}
$$

is the \emph{effective graph resistance} (or \emph{Kirchhoff index}) of the network, i.e. the sum of the resistances between all possible pairs of nodes in the graph (see, e.g., \cite{Klein1993}). $R$ reflects the overall transport capability of the network: the lower $R$, the better the network conducts flows. In particular, it has been shown that this index is able to catch the average vulnerability of a connection between a pair of nodes and, therefore, it is a suitable tool for assessing the ability of a network to well react when it is subject to failure and/or attack (see \cite{Ellens, WangP, CleCor}). \\

Effective resistances allow to give a specific definition of the centrality of a node in the network. Indeed, the \emph{best spreader} (or best connected)  node in the network is the node $i^\star$ that minimizes the quantity $\sum_{j=1}^{n}\omega_{i^\star j}=({\bf \Omega} \textbf{u})_{i^\star}$, i.e. the sum of all its resistance distances from any other node in the network. Since $L^{+}_{ii}$ equals the difference between the average resistance between node $i$ and all the other nodes in the network and the overall network mean resistance, then the best spreader node $i^\star$ is the one such that $L^{+}_{i^\star i^\star}\leq L^{+}_{jj}$ for any $j \neq i^\star$. Node $i^\star$ can be regarded as the best diffuser of a flow to the rest of the network, and, to some extent, it is the most influential with respect to a diffusion process inside the network, since it guarantees the highest flow toward other nodes (see \cite{Mieghem2017}). Best diffuser means that most of the information coming out from this node is absorbed by other nodes. If $L^{+}_{ii}$ is big, then most of this information comes back to node $i$ and doesn't reach other nodes. The reciprocal of $L^{+}_{ii}$ can then be regarded as a centrality measure of a node and it is called {\it vibrational centrality}. 


\section{Community detection based on communicability metrics}
\label{sec:6}

\subsection{The model}
\label{sec:6.1}
As discussed in the previous section, $\xi_{ij}=G_{ii}-2G_{ij}+G_{jj}$ and $\omega_{ij}=G^{v}_{ii}-2G^{v}_{ij}+G^{v}_{jj}$ represent the two metrics induced on the network by the Estrada communicability and the vibrational communicability, respectively. 
 
In an economical context, referring to the international trade network, they measure how well two countries, or companies, communicate in terms of commercial and trade exchanges. For instance, the attributes on nodes may be identified with the GDP and the currents along nodes with the total trade or money flow between two countries. 
Information on the network may be replaced by money flow. Therefore the quantity $\xi_{ij}$ of equation \ref{communicabilitydistance} accounts for the difference in the amount of money flow that returns to the nodes $i$ and $j$ and the amount of money flow exchanged between them. The bigger is $G_{ij}$, i.e. the money flow exchanged, the nearer are the nodes; the bigger are $G_{ii}$ or $G_{jj}$, i.e. the amount of money flow that comes back to the each node, the farther they are. 
A similar interpretation holds for $\omega_{ij}$. 
In a trade network $\omega_{ij}$ accounts for the difference between the mean resistance to export a given money flow from each country and the correlation between them. The bigger is $G^{v}_{ij}$, the more interconnected they are and the nearer they are in the resistance metric; the bigger are $G^{v}_{ii}$ and $G^{v}_{jj}$, the more isolated they are in the network and between them and the farther they are.

In light of these observations, we formulate our proposal\footnote{In what follows, we will refer to the communicability distance $\xi$, but similar arguments may be repeated identically for the resistance distance $\omega$}, considering as
members of the same cluster nodes whose mutual distance is below a given threshold $\xi_0$. Specifically, we construct a new community graph where the elements of the adjacency matrix $\textbf{M}=[m_{ij}]$ are given by:

$$
m_{ij}=
\left\{ 
\begin{array}{ll}
1 & \ {\rm if}\ \xi_{ij}\leq \xi_0 \\ 
0 & \ {\rm otherwise} \\ 
\end{array}
\right.
$$

with $\xi_0$ threshold distance such that $\xi_0\in [\xi_{\rm min}, \xi_{\rm max}]$, being $\xi_{\rm min}$ and $\xi_{\rm max}$ the minimum and the maximum distances between couples of nodes, respectively. In this way, clustered groups of nodes that strongly communicate emerge, in dependence of the threshold.  
If $\xi_0$ is high enough, all nodes in the network are at a mutual distance lower than the threshold and the whole network behaves like a unique community. As $\xi_0$ decreases, there will be nodes too far, such that to be considered disconnected and then members of different clusters, entailing the emergence of islands of connected nodes. 
Hence, the number of communities depends on the threshold, precisely it increases as $\xi_0$ decreases.

It is important to observe that with the proposed methodology we do not choose any {\sl a priori} optimal number of communities. Our approach is more in line with the classic Girvan-Newman approach \cite{Newman2004}. 
 
\noindent The optimal partition is determined according to an optimization problem whose objective function is based on the idea of cohesion between nodes. Specifically, since we deal with distances, following the approach for clustering in metric spaces proposed by \cite{chang2016}, we provide a cohesion measure $\gamma_{ij}$ between two nodes $i$ and $j$, as follows:
$$
\gamma_{ij}=\left(\bar{\xi}_{j}-\bar{\xi}\right)-\left(\xi_{ij}-\bar{\xi}_{i}\right)
$$
where $\bar{\xi}_{i}=\frac{1}{n-1}\sum_{k\neq i}\xi_{ik}$ is the average distance between $i$ and nodes other than $i$ and $\bar{\xi}$ is the average distance over the whole network.
Thus, ${\xi}_{ij}-\bar{\xi}_i$ represents the {\it relative distance} between nodes $i$ and $j$ and $\bar{\xi}_j-\bar{\xi}$ represents the {\it relative distance} from a random node to the node $j$.

Two nodes $i$ and $j$ are said to be cohesive (or incohesive) if $\gamma_{ij}\geq 0$ ($\gamma_{ij}\leq 0$). Notice that $\gamma_{ij} \geq 0$ yields ${\xi}_{ij}+\bar{\xi}\leq \bar{\xi}_{i}+\bar{\xi}_{j}$,  i.e., intuitively, two nodes are cohesive if they are close to each other and, on average, they are both far away from the other nodes. In other words, $\gamma_{ij}$ can be interpreted as the gain (when positive) or the cost (when negative) related to the grouping of nodes $i$ and $j$ in the same cluster of a given partition.

We assume to maximize an objective function that represents the global cohesion function based on the mutual relative distances between every pairs of nodes.
Therefore, we refer to a specific modularity index defined as

\begin{equation}\label{modu}
Q=\sum_{i,j}\gamma_{ij}x_{ij}
\end{equation}

where $x_{ij}$ is a binary variable equal to $1$ if two nodes are in the same cluster and $0$ otherwise and $\gamma_{ij}$ is the cohesion measure between nodes $i$ and $j$. 
It is worth to notice that when the partition is made up of a unique community, equal to the entire network, $x_{ij}=1\ \forall i,j$. In this case\footnote{Notice that $\bar{\xi}=\frac{1}{n}\sum_{i}\bar{\xi}_i=\frac{1}{n(n-1)}\sum_{i,j}\xi_{ij}$} 

\begin{align*}
Q=\sum_{i,j}\gamma_{ij}&=
\sum_{i,j}\left( \bar{\xi}_{j}+\bar{\xi}_{i}-\bar{\xi}-{\xi}_{ij}\right) \\
&=n\sum_{j}\bar{\xi}_{j}+n\sum_{i}\bar{\xi}_{i}-n^2\bar{\xi}-\sum_{i,j}{\xi}_{ij}\\
&=2n^2\bar{\xi}-n^2\bar{\xi}-n(n-1)\bar{\xi}\\
&=n\bar{\xi}.
\end{align*}

On the other hand, when the partition consists of $n$ isolated nodes, $x_{ij}=0$ $\forall i \neq j$ then 
$$
Q=\sum_i\gamma_{ii}=\sum_{i}(\bar{\xi}_{i}-\bar{\xi})-({\xi}_{ii}-\bar{\xi}_{i})=2\sum_{i}\bar{\xi}_{i}-n\bar{\xi}=n\bar{\xi}.
$$ 
Thus, in these two extreme cases, $Q$ provides the same value $n\bar{\xi}$. 

\subsection{An illustrative example}
\label{sec:6.2}
\begin{figure}[h]
	\includegraphics[width=.8\linewidth]{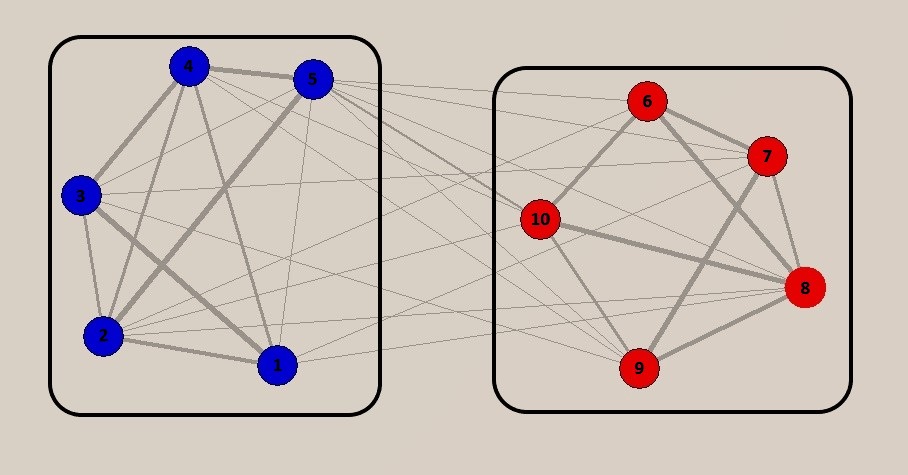}
	\centering
	\caption{A weighted undirected network with 10 nodes and 32 edges. Edges weights have been randomly sampled with replacement from integers between $1$ and $6$. The thickness of edges is proportional to the weights. Nodes of two relevant communities are highlighted in blue and red.
	}
	\label{fig:1}      
\end{figure}

We start by testing our methodology on a simple example. Let us consider the weighted undirected network displayed in Figure \ref{fig:1}. The network has $10$ nodes and $32$ edges. The thickness of links is proportional to weights. The network allows to easily identify two natural communities, which are highlighted by the two closed lines containing nodes $1$ to $5$ (on the left) and nodes $6$ to $10$ (on the right).

\begin{figure}[h]
	\includegraphics[width=.8\linewidth]{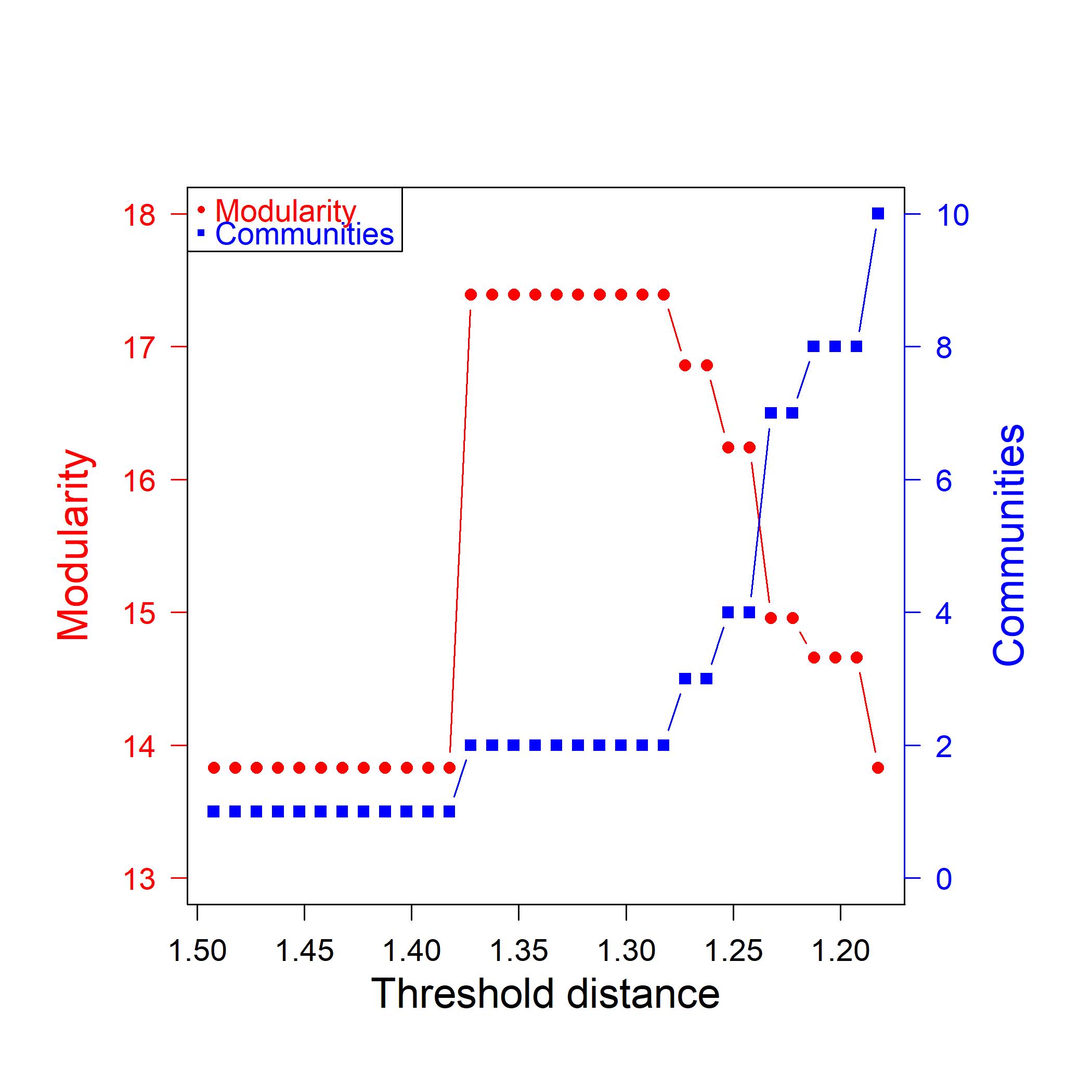}
	\centering
	\caption{Modularity $Q$ of the partition computed according to formula (\ref{modu}) and number of components (on the secondary scale) for different threshold values. The communicability distance has been used for the identification of the communities.}
	\label{fig:2}      
\end{figure}

\begin{figure}[h]
	\includegraphics[width=.8\linewidth]{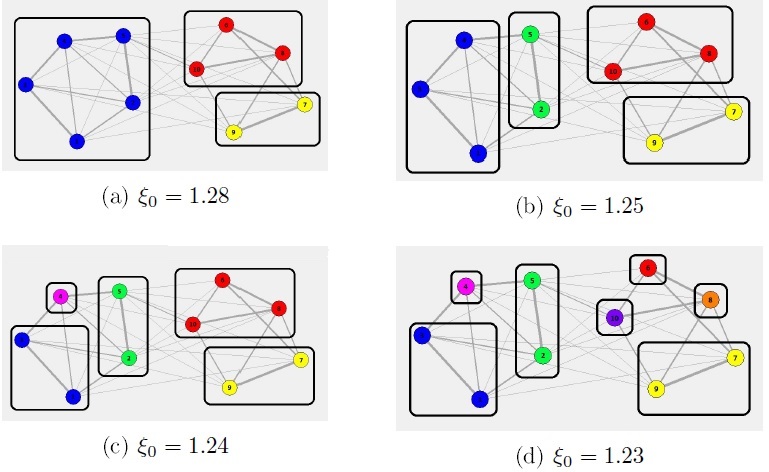}
	\centering
	\caption{Community structure at different thresholds.}
	\label{fig:3}      
\end{figure}

We compute the Estrada communicability matrix $\textbf{G}$, then we get the communicability distance matrix $\bf \Xi$. The nearest nodes are $1$ and $3$ with a communicability distance equal to $\xi_{\rm min}=\xi_{13}=1.18$ and farthest nodes are $3$ and $6$ with a communicability distance equal to $\xi_{\rm max}=\xi_{36}=1.49$. Figure \ref{fig:2} summarizes the number of communities identified at different thresholds. The blue line represents the number of communities while the red line represents the value of modularity of the corresponding partition. When the threshold is greater than or equal to $\xi_{0}=1.38$ all nodes are connected and the network is partitioned in a single community, with modularity $Q=n\bar{\xi}$. As the threshold decreases below $1.38$, the network begins to split into disconnected components. When the threshold becomes lower than the minimum distance, the network is partitioned into ten communities and each node belongs to a different community. The best partition according to the maximum modularity criterion splits the network into two clusters, which are easily identified with the two expected natural communities. The composition of the communities for alternative thresholds is reported in Figure \ref{fig:3}. It is noticeable that, lowering the threshold, the procedure allows to disentangle tightest relationships. For instance, when $\xi_{0}=1.23$ only nodes connected by edges with highest weights are kept in the same community.

\begin{figure}[h]
	\includegraphics[width=.8\linewidth]{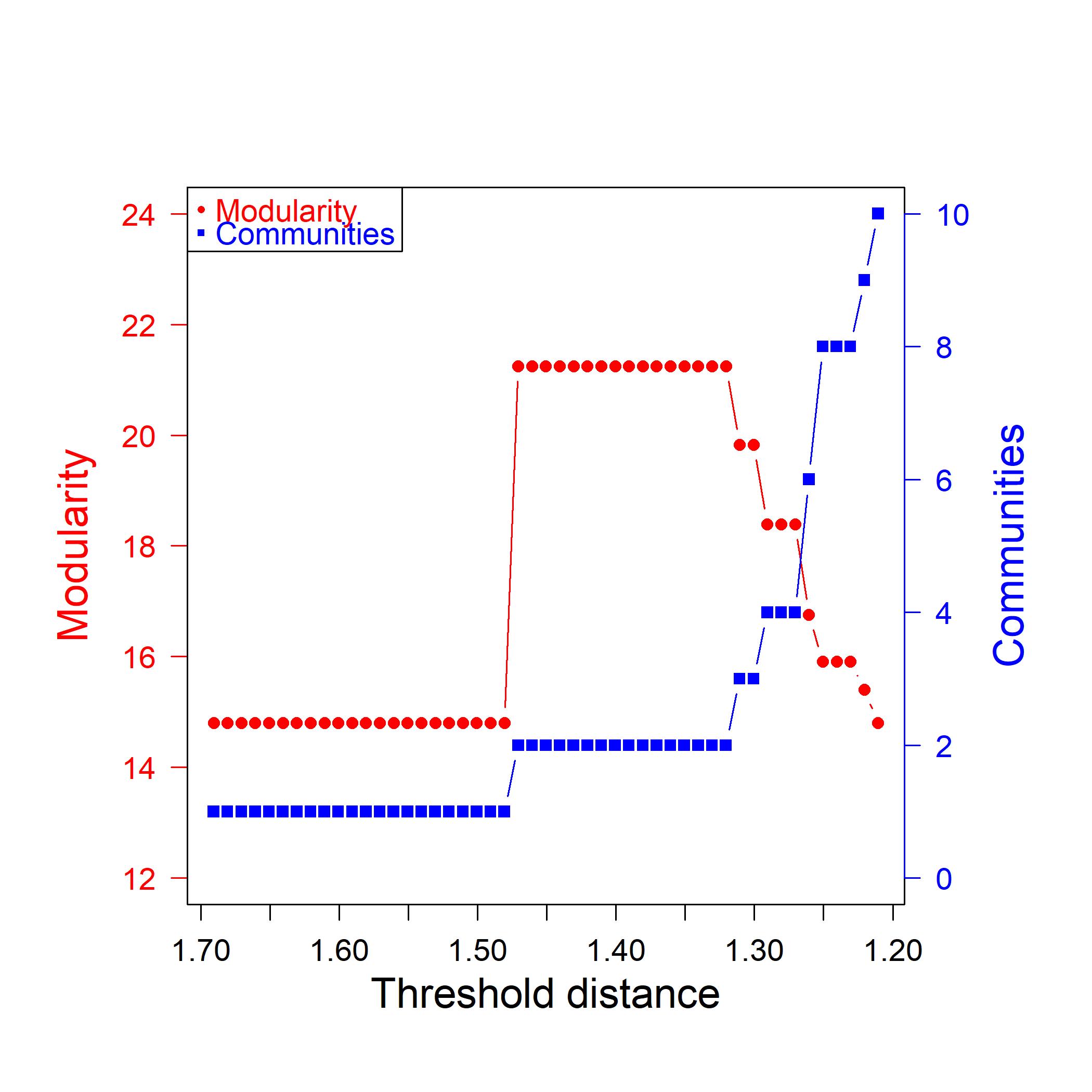}
	\centering
	\caption{Modularity of the partition and number of components (on the secondary scale) for different thresholds. The resistance distance has been used for the identification of the communities.}
	\label{fig:4}      
\end{figure}

Similar results are derived \added{by} applying the procedure based on the vibrational communicability. 
The nearest nodes are $1$ and $3$ with a resistance distance equal to $\omega_{\rm min}=\omega_{13}=1.22$ and farthest nodes are $3$ and $8$ with a resistance distance equal to $\omega_{\rm max}=\omega_{38}=1.69$. Again if we move the threshold from the maximum distance to the minimum distance, we get an increasing number of communities from $1$, the whole network, to $10$, isolated nodes. The best partition according to the maximum modularity criterion splits the network into the two expected communities, as shown in Figure \ref{fig:4}.

\section{Application to the World Trade Network}
\label{sec:7}
In this Section, we apply the proposed model in order to detect relevant communities of countries in the WTN. As described before, the method aims at grouping strongly interacting countries by means of their mutual distances. Two alternative distance functions will be tested. On the one hand, we find clusters exploiting communicability distance. Therefore we detect how much two countries are close in the network considering all possible weighted walks connecting them. On the other hand, we select clusters by means of resistance distance. In this case countries are grouped together if they have a similar relevance in the network in terms of vibrational centralities as well as if they are correlated in terms of their expositions towards common countries. \\
We start with a general description of the dataset and the main characteristics of the WTN. Then, we briefly summarize the primary steps of the methodology, providing a pseudo-code of the algorithm. Finally, we report the results in terms of community structure with the related discussion.

\subsection{Dataset and main characteristics of the WTN}
\label{sec:7:1}
\begin{figure}[h]
	\centering
	\includegraphics[width=.70\linewidth]{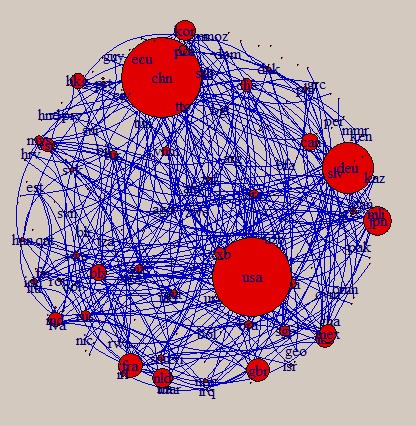}
	\caption{WTN based on 2016 data. Nodes are countries and 
		links are product trades between pair of countries. The size of the node is proportional to its strength.}
	\label{fig:5} 
\end{figure}

We refer to the World Trade Data, available on the Observatory of Economic Complexity database\footnote{The Observatory of Economic Complexity (OEC) is the world's leading data visualization tool for international trade data. Data can be found at: https://atlas.media.mit.edu/en/}. The database has been developed by the Research and Expertise Center on the World Economy at a high level of product disaggregation and it is based on original data provided by the United Nations Statistical Division (UN Comtrade). In particular, a harmonization procedure, that reconciles the declarations of exporters and importers, enables to extend considerably the number of countries for which trade data are available, as compared to the original dataset. In this analysis, we refer to the last version published in 2017, based on the Harmonized Commodity Description and Coding System, and that provides aggregated bilateral values of exports for each couple of origin and destination countries, expressed in billion dollars. We focus on the aggregated data of last available year, namely, 2016.
Hence, we construct a weighted network where each node is a country and weighted links represent the amount of product traded between couple of countries (see Figure \ref{fig:5}). The mutually exchanged products between two countries are different in terms of entity, so that they can be better represented by oriented links from a country to another one. However, we observed a strict relation between in and out strength distribution with a Spearman correlation coefficient equal to $0.956$. Hence, countries are ranked in a very similar way in terms of in and out strength. Thus, we performe all the analysis assuming the network as undirected.

The network is characterized by 221 nodes and 26197 links. The network is connected and its density is approximatively 0.539: on average, each country has trades with more than a half of the entire network. However, the network is not regular and is far from being complete or, in other words, most countries do not trade with all the others, but they rather select their partners. Furthermore, main trade flows tend to be concentrated in a specific sub-group of countries and a small percentage of the total number of flows accounts for a disproportionally large share of world trade. For instance, the top 10 countries export more than 50\% of the total flow. The maximum weight corresponds to the channel from China to USA and its value amounts to $436$ billion dollars. Minimum, non null, weights are involved in the trade between a number of very small countries, far from each others, and they are approximatively around $1$ thousand dollars. 
Finally, we expect that several countries trade with their geographical neighbours so that we investigate the correlation between flows and geographical distance of countries. 
We computed the Spearman rank correlation between link weights (i.e. monetary flows between countries in the network) and the great circle distance between capital cities in kilometers. We obtained a rank correlation of $-0.27$, that confirms a little preference for trading with physical neighbours. However, as stressed before, our aim is to go beyond immediate neighbours by means of both communicability and resistance distances.

\subsection{Summary of the methodology}
\label{sec:7:2}
In this section we summarize by means of a pseudo-code the main steps of the methodology we are proposing. 
The code has been written taking into account the communicability distance matrix $\bf \Xi$, but the same procedure can be easily applied by considering the resistance matrix $\bf \Omega$.

	\begin{enumerate}
		\item let ${\mathscr G}$ be the original directed weighted network with $n$ nodes and weighted adjacency matrix $\textbf{W}$;
		\item build the undirected weighted network ${\mathscr G}_1$ with a symmetric adjacency matrix defined as $\textbf{W}_{1}=\frac{1}{2}(\textbf{W}+\textbf{W}^T)$;
		\item build the undirected weighted network ${\mathscr G}_2$ with normalised weighted adjacency matrix $\textbf{W}_{2}= \textbf{S}^{-1/2}\textbf{W}_{1}\textbf{S}^{-1/2}$, where $\textbf{S}$ is the diagonal matrix of the strengths of the network ${\mathscr G}_1$; 
		\item construct the distance matrix ${\bf \Xi}=\textbf{g}\textbf{u}^T-2\textbf{G}+\textbf{u}\textbf{g}^T$ based on the communicability matrix $\bf G$;  
		\item define the threshold interval $[\xi_{\rm min}, \xi_{\rm max}]$, where $\xi_{\rm min}$ and $\xi_{\rm max}$ represent the minimum and the maximum communicability distances between couples of nodes, respectively and set $\xi_h=\xi_{\rm min}$, with $h=0$;
		\item define a $n \times n$ matrix $\textbf{M}_{h}=[m_{ij}]$ such that 
		$$
		m_{ij}=
		\left\{ 
		\begin{array}{ll}
		1 & \ {\rm if}\ \xi_{ij}\leq \xi_h\  {\rm and }\  i \neq j \\
		0 & \ {\rm otherwise} \\ 
		\end{array};
		\right.
		$$
		\item build the undirected unweighted network ${\mathscr G}_{3,h}$ from the binary adjacency matrix $\textbf{M}_{h}$;
		\item select the partition $P_{h}$ given by the components of the network ${\mathscr G}_{3,h}$;
		\item compute the modularity $Q=\sum_{i,j}\gamma_{ij}x_{ij}$ of the network ${\mathscr G}_{2,h}$ with respect to the partition $P_{h}$;
		\item set the number of iterations $r$, compute $k=\frac{\xi_{\rm max}-\xi_{\rm min}}{r}$, set $\xi_h=\xi_{h-1} + k$ and $h=h+1$ and repeat steps 6-9 until $\xi_h \leq \xi_{\rm max}$;
		\item select the optimal partition $P^{\star}_{h}$  as the partition $P_{h}$ that provides the maximum modularity $Q$.
\end{enumerate}

We stress some key points of the presented methodology. We aim at clustering countries on the basis of a specific distance. The two distances we have chosen highlight relationships of a different nature between countries and the different community structure emerging will support this fact. Varying the threshold we can disentangle the role of very tight relationships between couples of countries. Of course, reducing the threshold distance a great number of isolated nodes may appear. They are typically very small countries whose trade volume is very low and whose commercial partners are few. They play a marginal role in the WTN and they do not affect in a significant way the structure of the network in terms of relevant communities. This is the reason why we will focus our attention on the main communities that are produced by our methodology.

\subsection{Results}
\label{sec:7:3}

\subsubsection{Results in terms of communicability metric}
We initially applied the methodology described in Section \ref{sec:7:2} by using the communicability distance. The rationale for using the communicability metric on the WTN is the following. Two countries share a total volume of trade because they exchange a given set of products, of any kind. But they can be linked even if they don't exchange each other a given product, that is there is no direct flow of such product between them. A higher order exchange may occur between them. For instance, a country $A$ exports some raw materials - let's say, iron - to a country $B$; country $B$ produces mechanical parts from iron and exports them to country $C$. $A$ and $C$ communicate via a higher order walk and they depend on each other even if the two countries are not neighbours in the network. Indeed, communicability takes into account precisely all possible weighted walks between two nodes. 

\begin{figure}[h]
	\centering 
	\includegraphics[width=.80\linewidth]{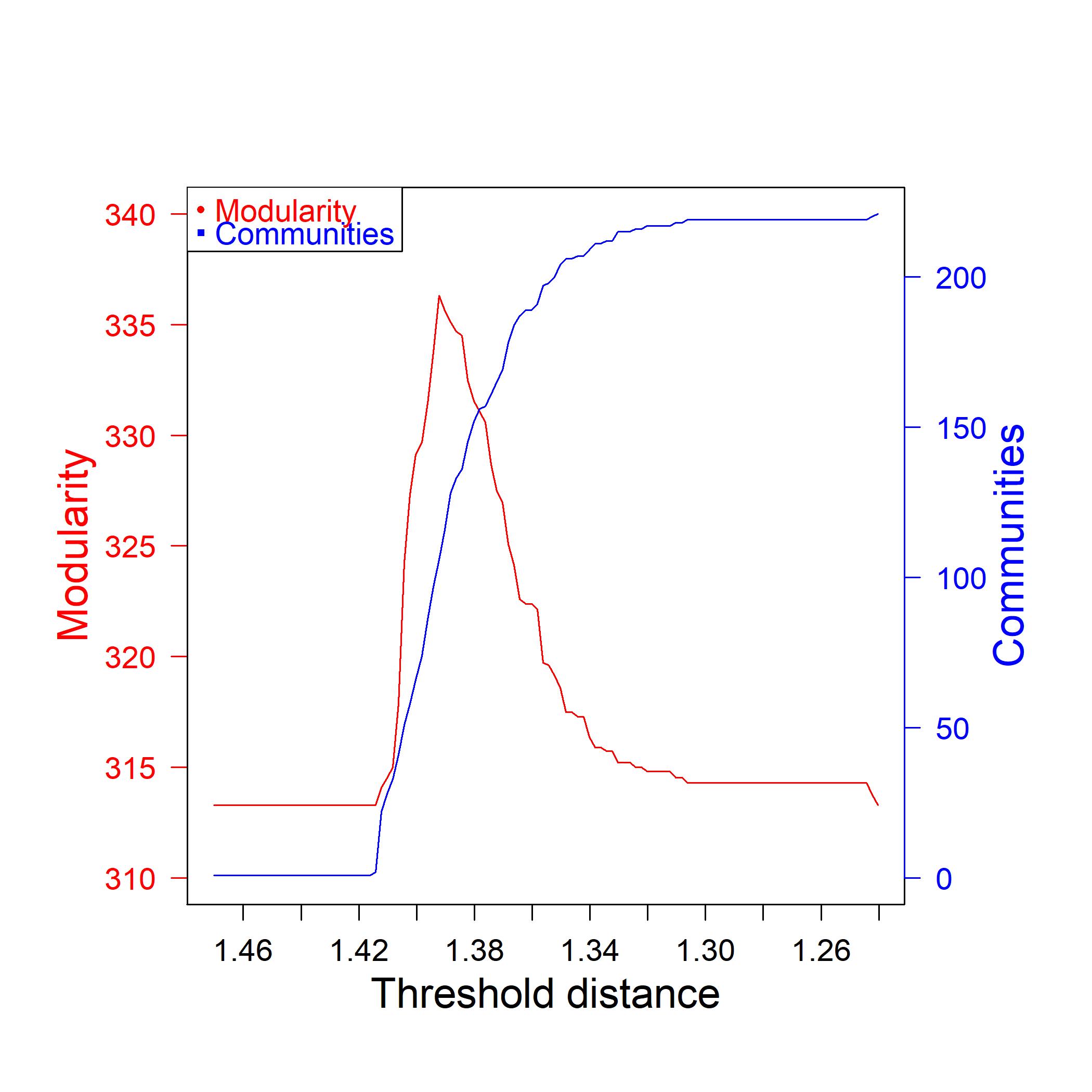}
	\caption{Modularity $Q$ (red line) and number of communities (blue line) as functions of the threshold communicability distance $\xi_h$}.
	\label{fig:6}
\end{figure}

Therefore, we calculate the communicability matrix $\textbf{G}$ on the normalised network ${\mathscr G}_2$ and the corresponding communicability distance matrix $\bf \Xi$. Using this metric, we find that the nearest countries are USA and Canada with a distance $\xi_{\rm min}=1.242$ and the farthest countries are USA and Seychelles Islands with a distance $\xi_{\rm max}=1.470$. For each value of the threshold distance between minimum and maximum, we look at the corresponding partition in communities. In Figure \ref{fig:6}, we plot the value of modularity $Q$ (in red) and the number of communities (in blue), counting each isolated node as an independent one. Both values are expressed as functions of the threshold $\xi_h$. The maximum of modularity is reached at a threshold distance $\xi_h=1.392$. It corresponds to $106$ communities, among which we have $87$ isolated nodes. Hence, we observe $19$ significant communities other than isolated nodes.

\begin{figure}[h]
	\centering
	\includegraphics[width=1\linewidth]{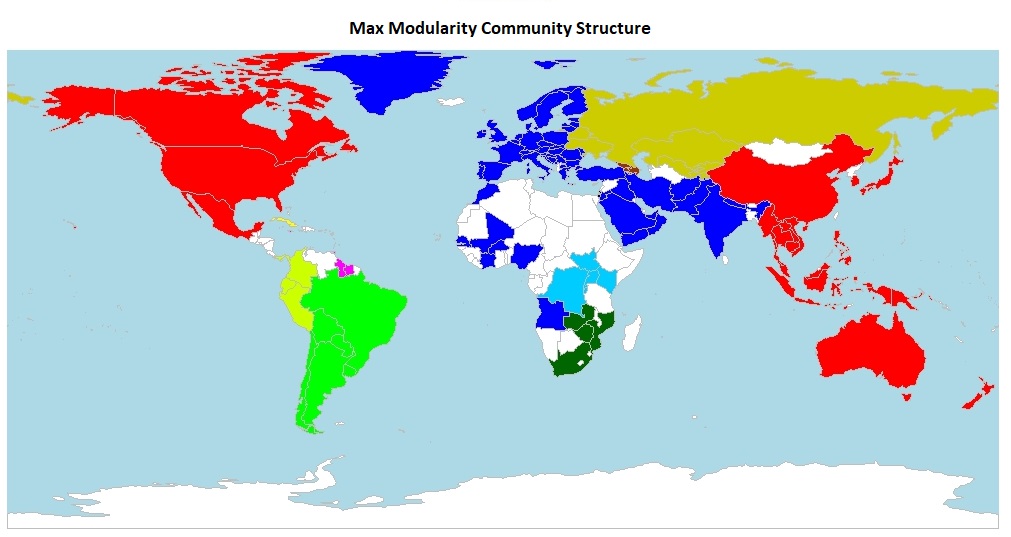}
	\caption{Optimal community structure based on communicability distance. Results have been derived by means of a maximum modularity criterion, with threshold $\xi_h=1.39$ (Max Mod Partition). Isolated nodes appear in white.}
	\label{fig:7}
\end{figure}

We display in Figure \ref{fig:7} communities in the optimal partition and we list in Table \ref{tab:2} the countries belonging to the 
ten biggest communities in terms of 
numerousness (excluding the one that groups isolated nodes).

Going deeper into the composition of the communities, the biggest one (see community 1 in blue) includes almost all continental European countries, with Great Britain and Ireland. This community acts on the screen of the global network as single player. It is worth pointing out also the presence of Morocco, confirming positive effects of bilateral trade agreements (see, e.g., \cite{Berkum2013}). We also notice the presence of South Asian countries that are economically linked togheter by the South Asian Association for Regional Cooperation. Presence of these countries in the community is also an effect of the bilateral foreign relations between the European Union (EU) and the Association of Southeast Asian Nations (ASEAN). The partnership between the EU and ASEAN dates back to 1972 when the EU countries became ASEAN's first formal dialogue partner. Finally, to the same community belongs african countries that are characterized by close economic and cultural ties to european countries, in particular to France (see, for instance, Ivory Coast, Burkina Faso, Angola, Senegal). \\
Opposed to this community, we see the second largest community (see, community 2 in red) which sees United States and China as main actors. This means that in Europe there are preferential channels of internal exchanges, whereas, outside Europe, most communication channels seem to be polarized around the exchange channel between China and the US and all their satellites countries. Moreover, we can recognize other well-identified and coherent communities. \\
Furthermore, it is interesting the decomposition of post-Soviet States. While Baltic and Eastern Europe States (except for Ukraine) have main partners in European countries, Central Asian countries have Russia as their leading trade and economic partner (see community 3).  Although a positive trade balance and a priority of Russian government of an increasing participation in  the  economic  relations  of Asia-pacific region (see \cite{Kuznetsova2016}), at moment, results show preferential channels with border countries. Transcaucasia is instead detected as a separate community (see community 10).\\
Except for Mexico, characterized by strong ties with United States, the Latin American and the Caribbean Economic System is decomposed into four relevant communities  (see communities 4, 6, 7 and 8). In particular, it is noticeable community 4 developed on the basis of the South Common Market, namely the so-called MERCOSUR. Mercosur's purpose is to promote free trade and the fluid movement of goods, people, and currency in south America. Since its foundation, Mercosur's functions have been updated and amended many times; it currently confines itself to a customs union, in which there is free intra-zone trade and a common trade policy between member countries. In 2019, the Mercosur had generated a nominal gross domestic product (GDP) of around 4.6 trillion US dollars, reaching the fifth economy of the world.  \\
Finally, significant blocks are also observed in central and south Africa (communities 5 and 9, respectively), polarized around  Democratic Republic of the Congo and Republic of South Africa.

\begin{table}[h]
	\centering{}
	\begin{tabular}{|lll|}
		\hline \hline
		& \bf Size & \bf Members \tabularnewline
		{\color{blue} \bf Community 1}  & \bf 54 & {\color{blue} \small \bf AFG AGO ARE AUT BFA BGR BHR BIH BLX}\tabularnewline
		&    & {\color{blue} \small \bf CHE CIV CYP CZE DEU DNK ESP EST FIN}\tabularnewline
		&    & {\color{blue} \small \bf  FRA GBR GRC GRL HRV HUN IND IRL IRN}\tabularnewline
		&    & {\color{blue} \small \bf  IRQ ITA JOR LTU LVA MAR MDA MKD MLI}\tabularnewline
		&    & {\color{blue} \small \bf   MNE NGA NLD NOR NPL OMN PAK POL PRT}\tabularnewline
		&    & {\color{blue} \small \bf   ROU SAU SEN SRB SVK SVN SWE TUR YEM}\tabularnewline
		\hline
		{\color{red} \bf Community 2} & \bf 21 & {\color{red} \small \bf AUS CAN CHN HKG IDN JPN KHM KOR LAO}\tabularnewline
		&    & {\color{red} \small \bf MEX MHL MMR MYS NZL PHL PNG SGP THA }\tabularnewline
		&    & {\color{red} \small \bf USA VNM XXB  }\tabularnewline
		\hline                             
		{\color{olive} \bf Community 3} & \bf 7 & {\color{olive} \small \bf BLR KAZ KGZ RUS TJK UKR UZB}  \tabularnewline
		\hline
		{\color{green} \bf Community 4}& \bf 6 & {\color{green} \small \bf ARG BOL BRA CHL PRY URY}	 \tabularnewline
		\hline
		{\color{cyan} \bf Community 5}& \bf 6 & {\color{cyan} \small \bf BDI COD KEN RWA SSD UGA} \tabularnewline
		\hline
		{\color{lime} \bf Community 6}& \bf 5 & {\color{lime} \small \bf BES COL ECU PAN PER} \tabularnewline
		 \hline
		{\color{yellow} \bf Community 7}& \bf 5 & {\color{yellow} \small \bf CRI GTM HND NIC SLV} \tabularnewline
		\hline
		{\color{magenta} \bf Community 8}& \bf 4 & {\color{magenta} \small \bf GUY JAM SUR TTO} \tabularnewline
		\hline
		{\color{teal} \bf Community 9}& \bf 4 & {\color{teal} \small \bf MOZ ZAF ZMB ZWE} \tabularnewline
		\hline
		{\color{brown} \bf Community 10}& \bf 3 & {\color{brown} \small \bf ARM AZE GEO} \tabularnewline
		\hline\hline
	\end{tabular}\caption{Members of the top ten communities in terms of number of countries.}
	\label{tab:2}
\end{table}

\begin{figure}[h]
			\centering
			\includegraphics[width=1\linewidth]{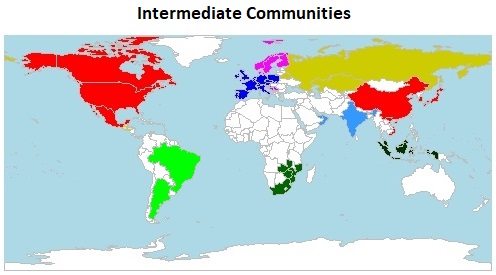}
			\caption{Intermediate Connected Community Structure - $\xi_h=1.37$}
			\label{fig:8}
		\end{figure}

\begin{figure}[h]
			\centering
			\includegraphics[width=1\linewidth]{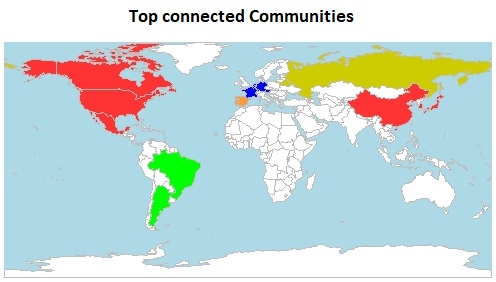}
			\caption{Top Connected Community Structure - $\xi_h=1.35$}
			\label{fig:9}
		\end{figure}

If we reduce the threshold, we let very strong channels of communication between countries emerge. For instance, Figures \ref{fig:8} and \ref{fig:9} show the community structure lowering the threshold distance (equal to $\xi_h=1.37$ and $\xi_h=1.35$, respectively). Moving from $1.39$ to $1.37$ some loose connections are lost (see Figure \ref{fig:8}). Scandinavia and the Nordic Region split up from community 1 creating a separate cluster together. The South East Asian and former Yugoslavia appear as separate communities characterized only by most relevant partnerships, Australia goes out from community 2, and the strong community in the South of Africa loses some country. Furthermore, in South America, only the relation between Brazil and Argentina survives. This result is in line with the fact that the strategic relationship between Argentina and Brazil is considered to be at the highest point in history: Brazil accounts indeed for Argentina's largest export and import market. \\
Reducing further the threshold to $1.35$, only the most closely interrelated communities survive. The strongest community counts now, among its members, all North America, Mexico, China and Japan (in red in Figure \ref{fig:9}). In Europe two communities are saved. On the one hand, the relation between Spain and Portugal is preserved. On the other hand, a community emerged in central europe around the channel between France and Germany. Finally community 3 in Table \ref{tab:2}, including Russia and Central Asian countries, resists also when the threshold is lowered.
		
\begin{figure}[h]
			\centering 
			\includegraphics[width=.80\linewidth]{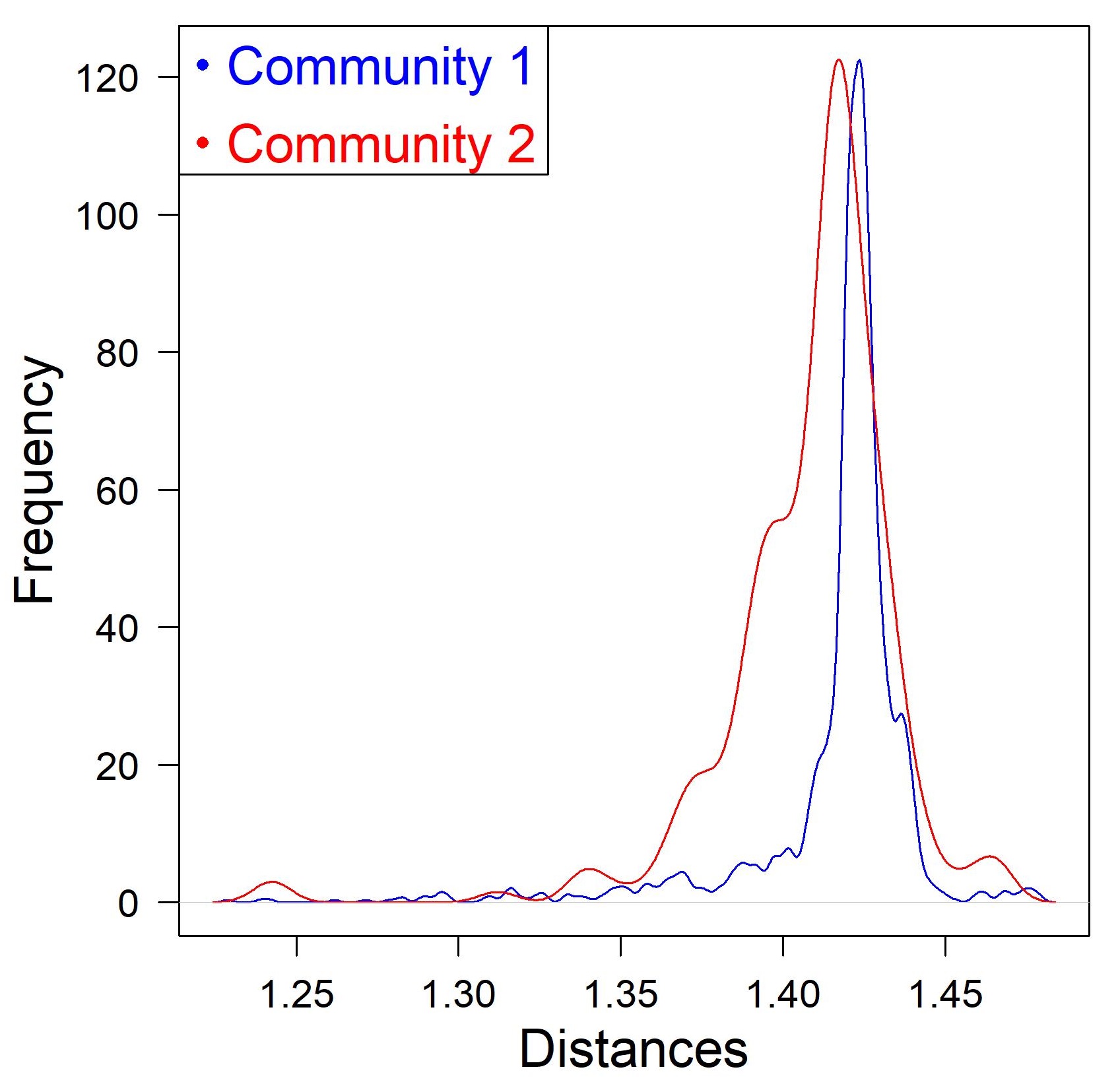}
			\caption{Distributions of Communicability Distances between countries of the some community. We display only the distributions related to the two main communities summarized in Table \ref{tab:2}}
			\label{fig:10}
		\end{figure}

		A significant feature of our approach is the fact that it allows to get deeper insight into the internal structure of each community and to give a measure of the mutual relationships between communities. Let us refer now to the clusters depicted in Figure \ref{fig:7} and detected with the maximum modularity criterion. In this regard, we display in Figure \ref{fig:10} the distributions of the communicability distances between pair of countries that belong to the same community. In particular, we compare the distributions for the first two relevant communities listed in Table \ref{tab:2}
		
		In fact, if we focus, for instance, on community 1 and 2, we can inspect and compare their internal structure by providing some synthetic indicators in Table \ref{tab:3}.
		From the analysis of Figure \ref{fig:10} and of the values shown in Table \ref{tab:3}, we can say that the community 2 (let's say, USA-China) shows slightly more intense interactions than community 1 (let's say, Europe) since in the former the average intracluster distance is slightly lower than in the latter. However, although the largest number of countries that belong to community 1, a more compact distribution is observed with a lower volatility. Trading interactions between countries in community 1 appear indeed somehow more homogeneous than between countries in community 2. This is partially related to the geographical distribution of the countries inside the two communities. We have indeed that community 2 can be interpreted as the aggregation of different blocks mainly developed around USA, China and Japan. \\
		Last column of Table \ref{tab:3} provides the same indicators computed on intercluster basis. This analysis allows to provide additional information in terms of heterogeneity in the group and between groups. It is worth pointing out the lower intercluster standard deviation. It means that couple of countries that belong to a different community has a similar distance between them.

		\begin{table}[h]
			\centering{}
			\begin{tabular}{||l|ll|l||}
				\hline 
				&  \multicolumn{2}{c|}{\bf Intracluster} & \bf Intercluster \tabularnewline
				&   Community 1 & Community 2 & Community 1 vs 2\tabularnewline
				\hline 
				Number of Nodes & 54 & 21 & --- \tabularnewline
				Mean Distance & 1.414 & 1.409 & 1.423 \tabularnewline
				Min Distance & 1.325 & 1.242  & 1.393\tabularnewline
				Closest Countries & NLD-BLX & USA-CAN & SAU-KOR   \tabularnewline
				Max Distance & 1.444 & 1.467 & 1.469\tabularnewline
				Furthest Countries & DEU-AFG & USA-LAO & USA-MNE  \tabularnewline
				Standard Deviation & 0.012 & 0.028 & 0.011 \tabularnewline
				\hline 
			\end{tabular}\caption{Intercluster and Intracluster characteristics of the distributions of communicability distances. Columns Community 1 and Community 2 refer to the intracluster properties of the two main detected communities, in terms of number of nodes. Last column reports the corresponding intercluster properties computed between the same two communities.}
			\label{tab:3}
		\end{table}

		It is noteworthy that additional insights can be provided by assessing the relevance of each country in the community. Indeed, communicability distance matrix provides a metric on the network and on each subnetwork, like a community. Therefore, we adapt the idea of closeness to our context, by providing the following communicability closeness to assess how effectively a node is supposed to spread trade flows through the network. Similarly to the definition of closeness, we define the \emph{communicability closeness} as:

		\begin{equation}
		C_{i}=\frac{1}{\sum_{j \in {\mathscr{C}}}\xi_{ij}}
		\label{Com}
		\end{equation}
				
		where the sum is over all the internal nodes of the cluster ${\mathscr{C}}$ to which the node $i$ belongs. \\
		
\begin{figure}[h]
			\centering 
			\includegraphics[width=.52\linewidth]{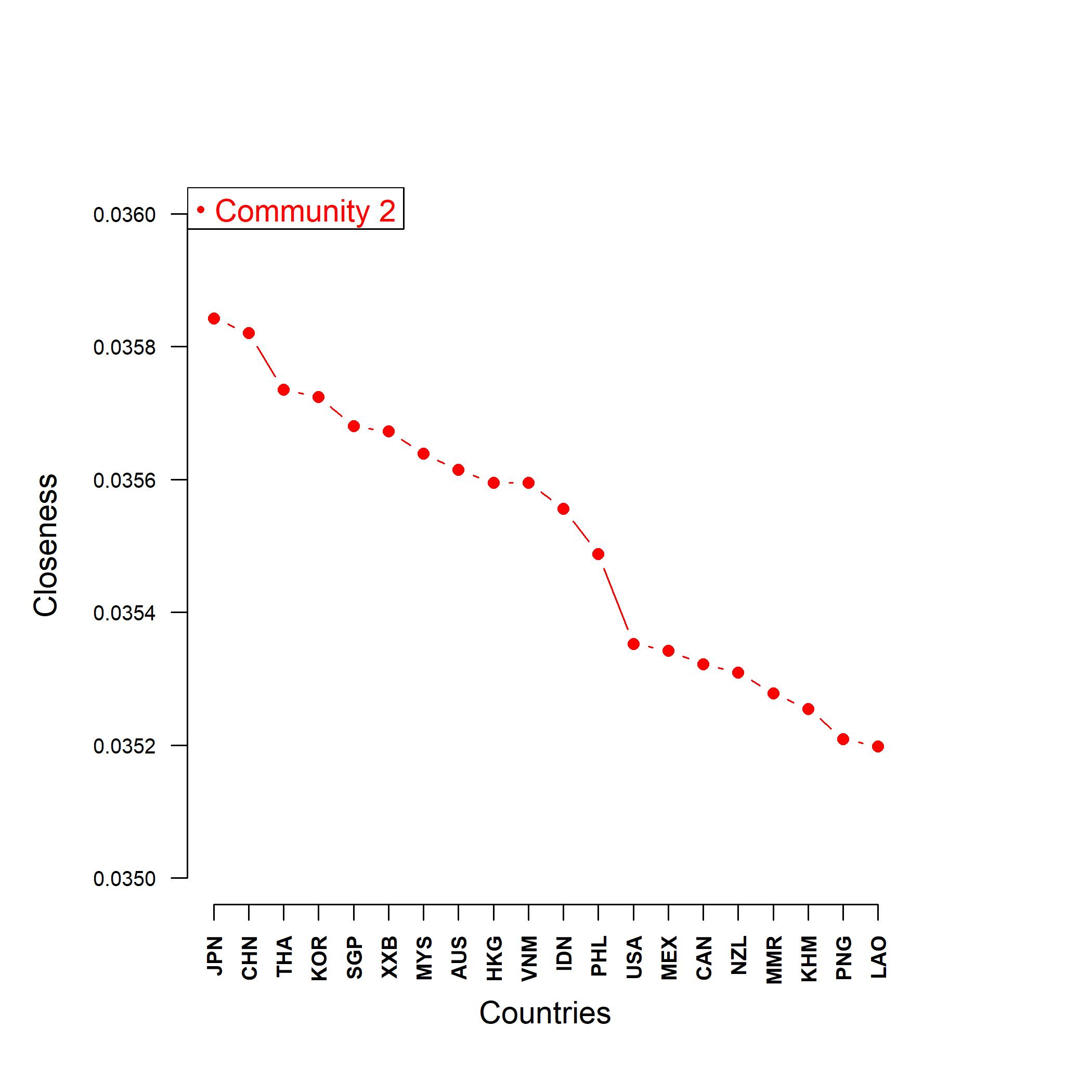}
            \includegraphics[width=.46\linewidth]{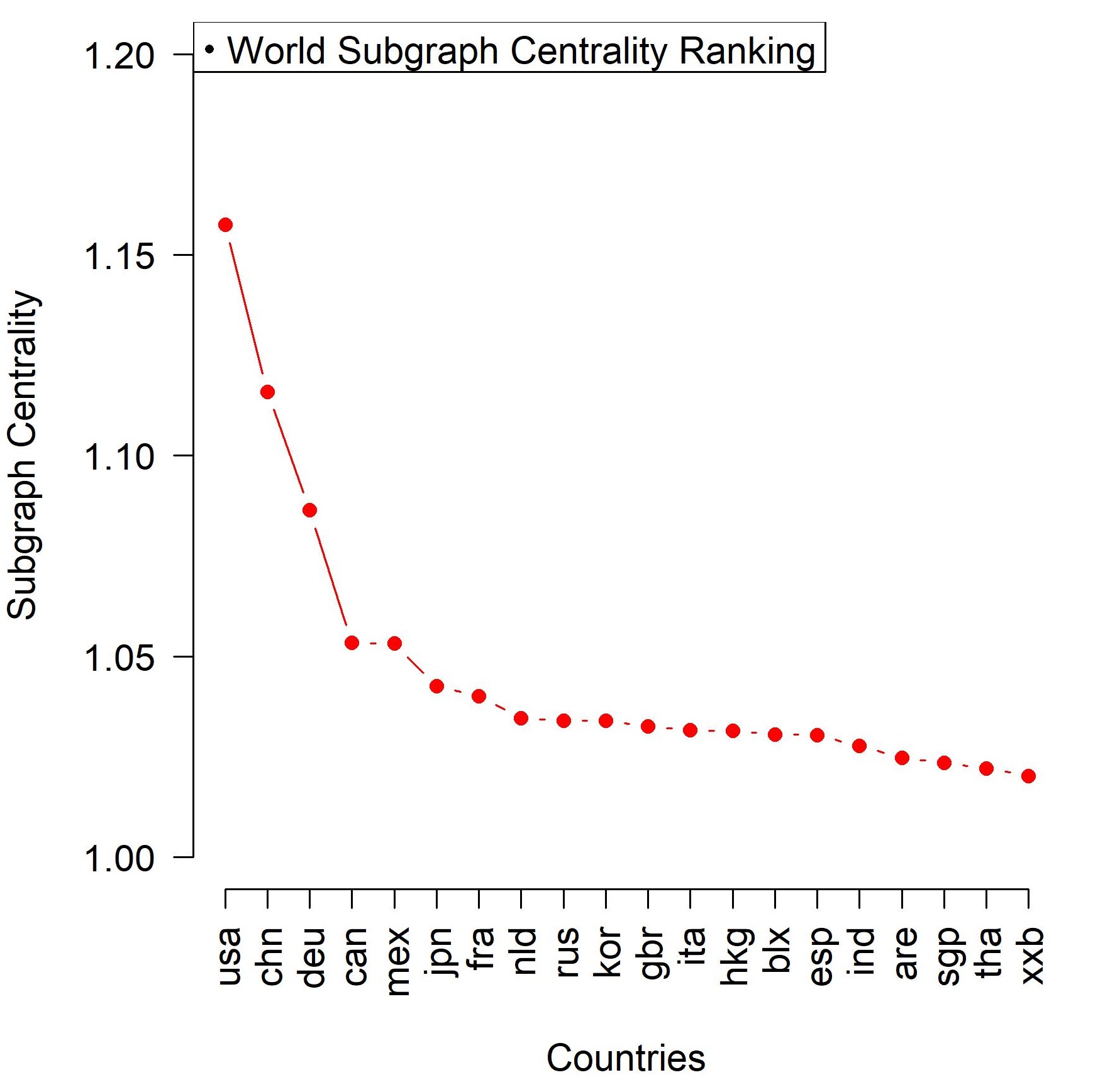}
			\caption{On the left-hand side, values of communicability closeness $C_{i}$ for the top 20 countries inside community 2; on the right-hand side, world top 20 countries according to subgraph centrality rankings.}
			\label{fig:12}
		\end{figure}

	To exemplify, we rank in Figure \ref{fig:12} (left-hand side) the top 20 countries of community 2 on the basis of values of $C_{i}$. It is worth to stress that the centre of this community is located in China, Japan and South Korea and not in the North American sub-community. The three Asian nations are nowadays major traders and their high-level economic cooperation has been strengthened also because of the speed-up of the negotiations on the trilateral Free Trade Agreement. The three parties unanimously agreed to further increase the level of trade and investment liberalization based on the consensus reached in the Regional Comprehensive Economic Partnership Agreement\footnote{See "Fifteenth Round of Negotiations on a Free Trade Agreement among Japan, China and the Republic of Korea", April 12, 2019, , Ministry of Foreign Affairs of Japan and  Free Trade Agreement (FTA) and Economic Partnership Agreement (EPA), 4 November 2019, Ministry of Foreign Affairs of Japan}. \\
				
		Moreover, it is interesting to see that most central country in a community has not necessarily the same relevance on the whole network. We have indeed that, in terms of subgraph centrality, when we deal with the whole network (see Figure \ref{fig:12}, right-hand side), USA appears as the key player followed by China and Germany. This ranking is inline with the top three countries provided by the World Trade Organizations, in terms of World's leading traders of goods and services \cite{WTO}.
		 
		Additionally, it is interesting to highlight that the relevance of countries reported in Figure \ref{fig:12} (right-hand side) is consistent with the Economic Complexity Index (ECI), introduced by \cite{Hausmann2014}. The ECI allows to rank countries in the WTN according to the diversification of their export flows, which reflects the amount of knowledge that drives their growth. The higher is the ECI, the more advanced and diversified is an economy. In particular, countries whose economic complexity is greater than expected (on the basis of their global income), tend to grow faster than rich countries with a low ECI. In this perspective, ECI represents a suitable tool for comparing countries in the WTN independently of their total output and it has been extensively validated as a relevant economic measure by showing its capability to predict future economic changes and to explain international differences in countries incomes.\\
		Although the network we analysed in the present work is based on the total normalised output and this fact prevents us from comparing directly their values with the ECI for a given country, there is a positive correlation between them. All the top 20 countries in Figure \ref{fig:12} (right-hand side) show a positive and high value of ECI. More specifically, they kept a high value of ECI during the years preceding the year to which the network refers (2016) and this can justify the high value in the aforementioned centrality measures. \\		
	 
\begin{figure}[h]
			\centering 
			\includegraphics[width=.49\linewidth]{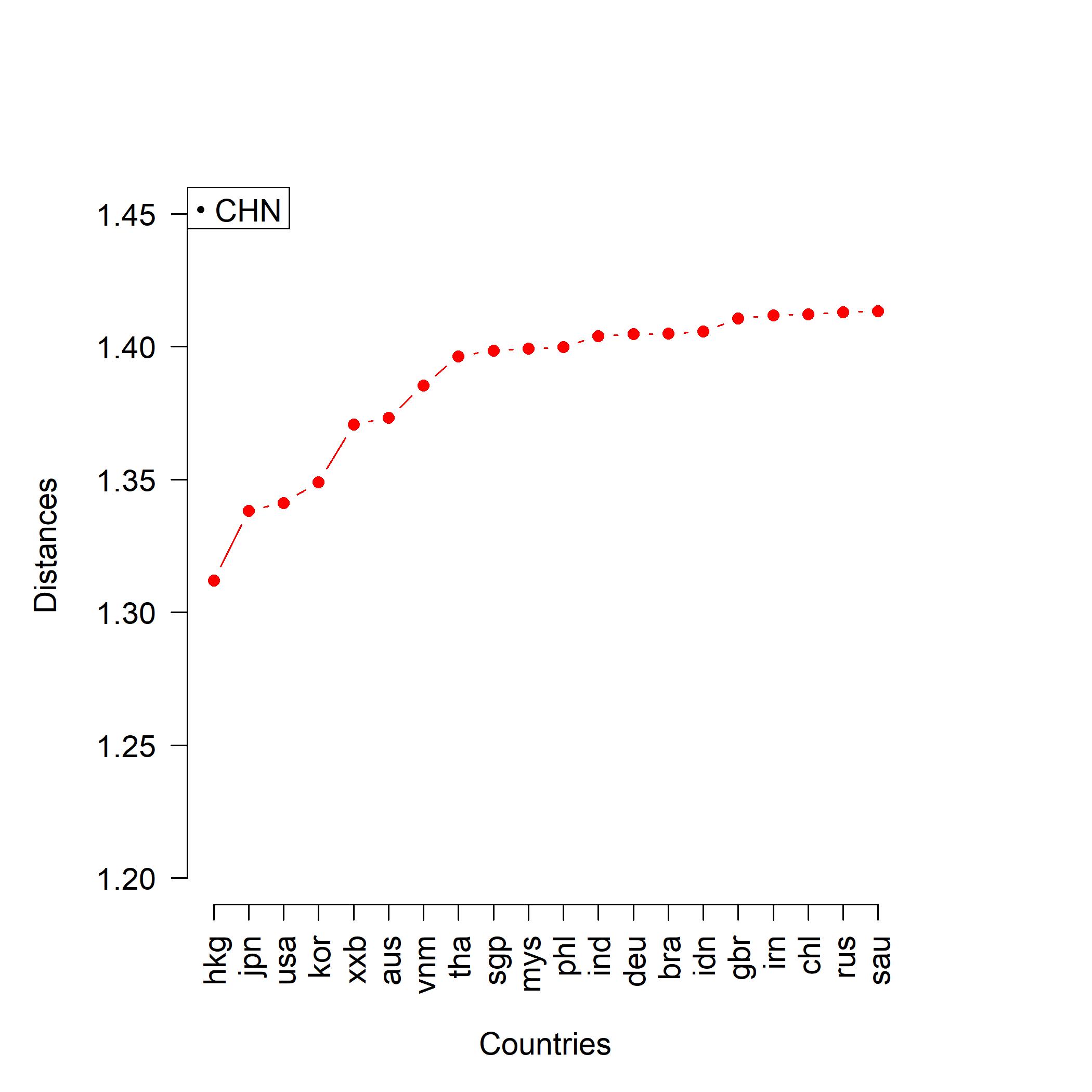}
			\includegraphics[width=.49\linewidth]{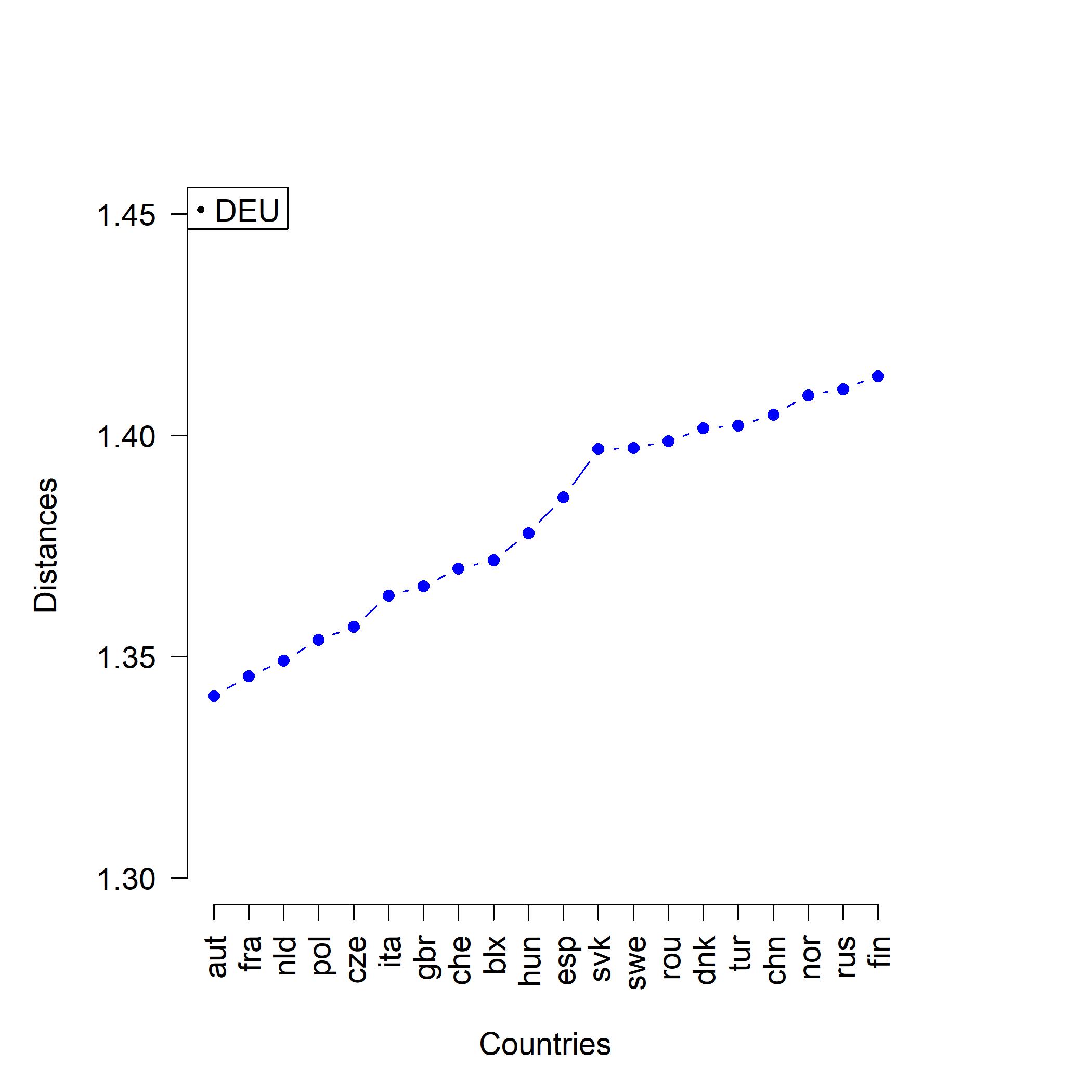}
			\caption{Top 20 nearest countries for China (left) and Germany (right)}
			\label{fig:14}
		\end{figure}

		Finally, from the point of view a single country, it is worth to look for the closest trade partners, that is the nearest nodes in terms of communicability distance. Figures \ref{fig:14} show the distance profiles for China and Germany, respectively. For instance, looking at Figure \ref{fig:14} (right-hand side), we can notice countries, as Austria, Poland, Czech Republic that are characterized by a condition of strong dependence on Germany, that is a major player in the network.
	    Similarly Figure \ref{fig:14} (left-hand side) shows how strong is the commercial relationship between China and Hong Kong, also as a result of the trade agreements between the two countries, like CEPA (Closer Economic Partnership Arrangement) aimed at eliminating duties on large categories of products. Indeed, it is well-known that, for the Chinese trade market, Hong Kong plays a crucial role since foreign companies use Hong Kong as a springboard to invest in China thanks to its infrastructure network that has no equal in the world, investor protection, transparent and efficient judicial system, legal certainty.

		\subsubsection{Results in terms of resistance metric}
		\label{sec:7.3.2}
		The methodology described in Section \ref{sec:7:2} has also been applied using the resistance distance $\omega$. In this case,	we consider the total trade of a given country as flow of the global wealth that has been produced during a year. Therefore, the Gross Domestic Product (GDP) is the attribute of interest on each node. In this regard, the effective resistance of an edge expresses how easily (or not) a unit flow moves from a country to another one, i.e. how easily two countries trade a unit of wealth, independently of its nature. It is noteworthy that, according to formula \ref{resistancedistance}, the resistance distance between a pair of countries depends on the values of the vibrational centralities of both countries (the more central these countries are in the network, the less is the resistance distance between them) 
		and on the value of their mutual correlation (the more correlated they are and again the less is their distance). 
		
\begin{figure}[h]
			\centering 
			\includegraphics[width=.80\linewidth]{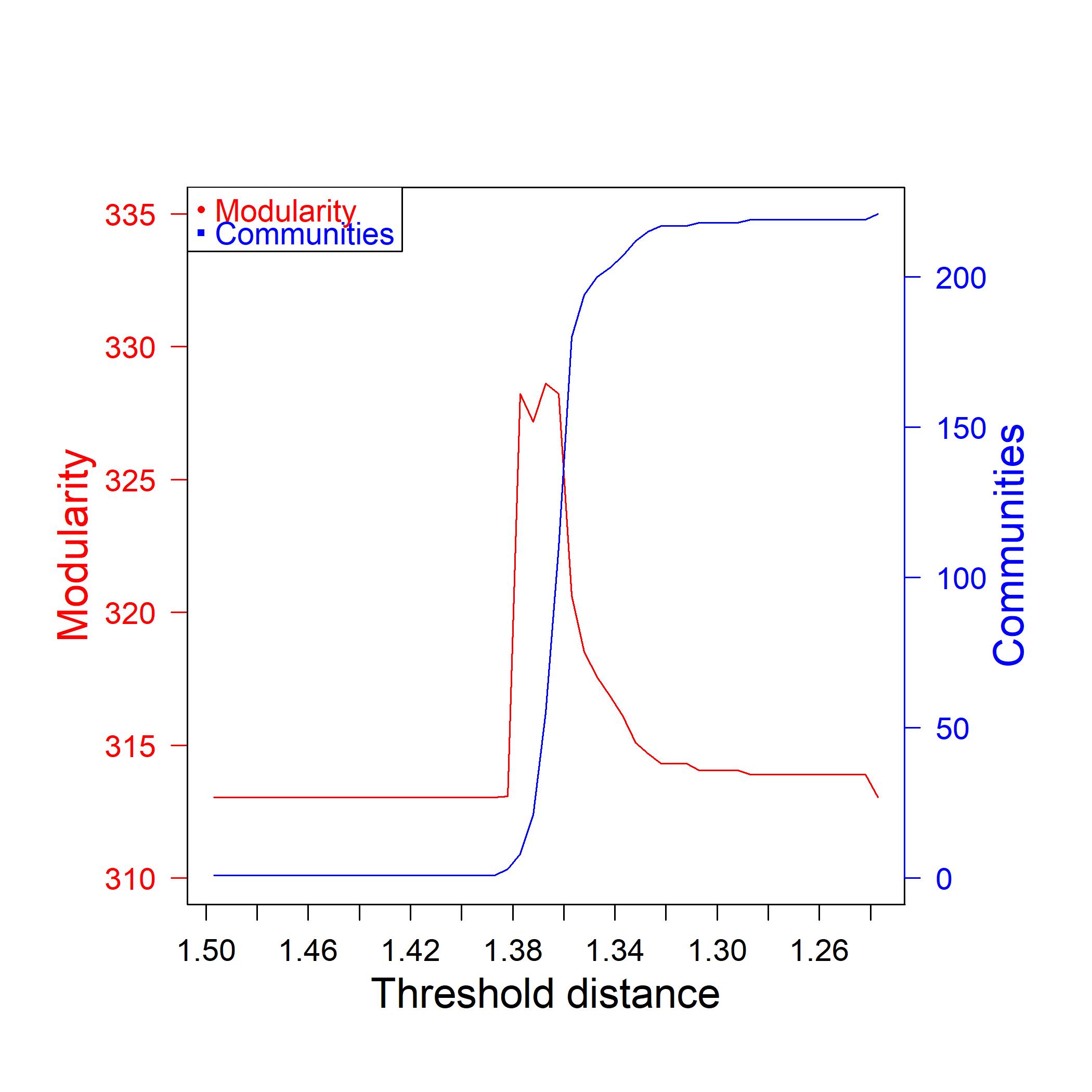}
			\caption{Modularity (red line) and number of communities (blue line) as functions of the threshold resistance distance. Maximum modularity is observed for  $\omega_h=1.365$}
			\label{fig:16}
		\end{figure}

\begin{figure}[h]
			\centering
			\includegraphics[width=1\linewidth]{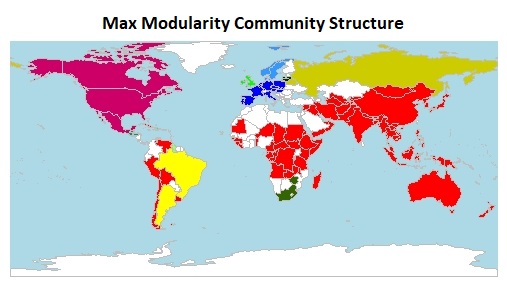}
			\caption{Communities detected by using the procedure based on the resistance matrix and considering the threshold $\omega_h$ that maximize the modularity.}
			\label{fig:17}
		\end{figure}

		Therefore, we construct the vibrational communicability matrix $\textbf{G}^v$ on the normalised network ${\mathscr G}_2$, and the corresponding resistance distance matrix $\bf \Omega$. Using this metric, we find that the nearest countries are, again, USA and Canada with a distance $\omega_{\rm min}=1.238$ and the farthest countries are USA and Germany with a distance $\omega_{\rm max}=1.497$. For each value of the threshold distance between minimum and maximum, we obtain the corresponding partition in communities. The maximum modularity partition corresponds to $15$ communities plus isolated nodes. In Figure \ref{fig:16}, we plot the value of modularity in red and the number of communities, counting each isolated node as an independent one, in blue as functions of the threshold $\omega_h$. The maximum modularity is reached at a threshold distance $\omega_h=1.365$. The main characteristic of this partition is the presence of a giant component of 127 nodes e 14 other components with few nodes. \\
				Main results in terms of geographical distribution are displayed in Figure \ref{fig:17} and, as in the previous Section, we summarize in Table \ref{tab:5} main composition of top communities in terms of number of constituents.
		
		With respect to results based on communicability, we have that the first community has a larger number of countries (equal to $127$). Additionally, the larger community includes again main asian and oceanian countries as well as several african countries. It is noteworthy that North America behaves as a separate cluster. This result is in line with the literature that emphasizes the interesting economic relation between Asia and Oceania. Several works showed that the Asia-Oceania community collapsed after China entered the WTO in 2001 and built strong trade relationships with other communities, especially with the external cores, (i.e. the United States and Germany). China then became regionally attractive and restored the Asia-Oceania community as the community leader after it gained a significant portion of trade globally (see, e.g., \cite{Zhu2014}).  \\
	   Significant differences are also observed for the European community (see community 2 in Table 			\ref{tab:5}). Norway and Sweden and Great Britain and Ireland provide indeed two separate groups with respect to main european economic groups.

It is worth pointing out that communities detected above represent groups of countries showing a positive correlation in their trade strength, whereas members of different clusters show a negative correlation. Being strongly anti-correlated means that when the total trade deficit of a country grows, the total trade surplus of a second country grows too. For instance, Japan and USA have been classified by the methodology in different communities. Indeed, in the literature, empirical analyses show a negative correlation coefficient between normalised trade strengths of these countries (see, e.g., \cite{kozmetsky2012} and \cite{delrio2017}). Similar arguments can be extended also to other pairs of countries. For instance,  Germany is negatively correlated with USA (see \cite{kozmetsky2012}) and show a high positive correlation with Belgium and France (see \cite{delrio2017}), that belong to the same community.
	
If we disentangle communities characterized by very tight relationships between countries, the results seem strictly related to the ECI index. We may expect that, if two countries communicate well, then their ECI's could be similar. That is, if their mutual distance is small, both in terms of communicability metric and resistance metric, then they display similar values of ECI. In fact, the existence of multiple channels of trade exchange between them would result in a similar diversification of their output. This means that countries inside each community (could) share homogeneous values of ECI. Concerning Table \ref{tab:5}, we notice small clusters whose components show homogeneous values of the ECI index. For instance, community 6 is formed by Russia (with an ECI of $0.855$ in 2016) and Belarus (with an ECI of $0.744$ in the same year). Similarly Canada ($1.084$), Mexico ($1.160$) and USA ($1.781$); Norway ($1.199$) and Sweden ($1.862$); UK ($1.549$) and Ireland ($1.409$); Brazil ($0.648$) and Argentina ($0.380$) that constitute communities 3, 4, 5 and 7, respectively.

		\begin{table}[h]
			\centering{}
			\begin{tabular}{|lll|}
				\hline \hline
				& \bf Size & \bf Members \tabularnewline
				{\color{red} \bf Community 1}  & \bf 127 & {\color{red} \small \bf AUS CHN HKG IDN IND IRN IRQ}\tabularnewline
				&    & {\color{red} \small \bf JPN KOR LAO PHL THA and others}\tabularnewline
				\hline
				{\color{blue} \bf Community 2} & \bf 11 & {\color{blue} \small \bf AUT BLX CZE DEU ESP FRA HUN }\tabularnewline
				&	 & {\color{blue} \small \bf ITA NLD POL PRT SVK}\tabularnewline
				\hline				    
				{\color{magenta} \bf Community 3} & \bf 3 & {\color{magenta} \small \bf CAN MEX USA}  \tabularnewline
				\hline
				{\color{cyan} \bf Community 4} & \bf 2 & {\color{cyan} \small \bf NOR SWE}  \tabularnewline  
				\hline
				{\color{green} \bf Community 5} & \bf 2 & {\color{green} \small \bf GBR IRL}  \tabularnewline
				\hline
				{\color{olive} \bf Community 6}& \bf 2 & {\color{olive} \small \bf BLR RUS} \tabularnewline
				\hline
				{\color{yellow} \bf Community 7}& \bf 2 & {\color{yellow} \small \bf ARG BRA}	 \tabularnewline
				
				\hline \hline
			\end{tabular}\caption{Members for the seven main communities in terms of number of countries. The optimal partition has been obtained by applying the procedure based on the maximum modularity and a threshold depending on the resistance distance.}
			\label{tab:5}
		\end{table}
		
		
		

		As in the previous Section, we explore main characteristics of two most relevant communities (see Table \ref{tab:6}), It is noticeable that, although the two groups show a very similar mean distance, European countries are characterized by a higher heterogeneity. Focusing on intercluster indicators, we notice also a lower similarity between the two communities with respect to Table \ref{tab:3} based on communicability.

		\begin{table}[h]
			\centering{}
			\begin{tabular}{||l|ll|l||}
					\hline 
					&  \multicolumn{2}{c|}{\bf Intracluster} & \bf Intercluster \tabularnewline
					&   Community 1 & Community 2 & Community 1 vs 2\tabularnewline
						\hline 
				Number of Nodes & 127 & 12 & --- \tabularnewline
				Mean Distance & 1.414 & 1.391  & 1.418 \tabularnewline
				Min Distance & 1.290 & 1.324  & 1.395\tabularnewline
				Closest Countries & CHN-HKG & AUT-DEU & SMR-DEU \tabularnewline
				Max Distance & 1.429 & 1.424  & 1.466 \tabularnewline
				Furthest Countries &  ARE-HKG & AUT-PRT & JPN-DEU\tabularnewline
				Standard Deviation & 0.008 & 0.026 & 0.010 \tabularnewline
				\hline 
			\end{tabular}\caption{Intercluster and intracluster characteristics of the distributions of resistance distances.  Columns Community 1 and Community 2 refer to the intracluster properties of the two main detected communities, in terms of number of nodes. Last column reports the corresponding intercluster properties computed between the same two communities.}
			\label{tab:6}
		\end{table}

\begin{figure}[h]
			\centering 
			\includegraphics[width=.80\linewidth]{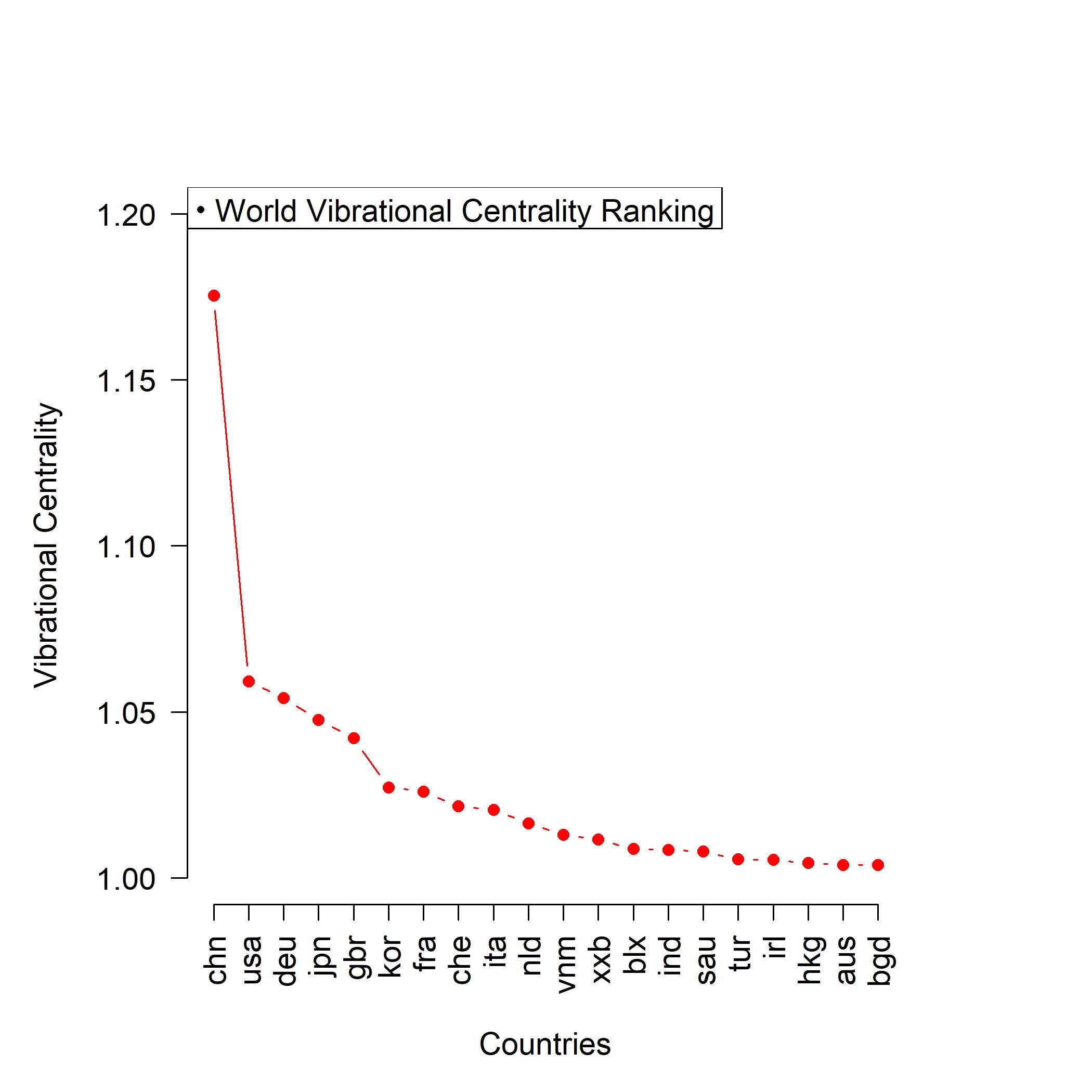}
			\caption{World top 20 countries according to vibrational centrality rankings}
			\label{fig:21}
		\end{figure}

		
		The relevance of a country can be now assessed in terms of vibrational centrality. To this end, we display in Figure \ref{fig:21}, the top 20 countries, calculated over the whole network. China, USA and Germany are again in the top 3, with China playing as the best spreader node. Also in this case, almost all the top 20 has a positive ECI. A comparison between Figures \ref{fig:12} and \ref{fig:21} confirms the different role played by USA and China in the global network. As confirmed by \cite{WTO}, USA is the leading commercial service provider and in such a way it is widespread well-integrated in the global market; on the other side, China plays the role of hub for goods and represents the leading merchandise trader and this gives to the country a very robust position which makes it less vulnerable to market turmoil. 
		
\begin{figure}[h]
			\centering
			\includegraphics[width=.49\linewidth]{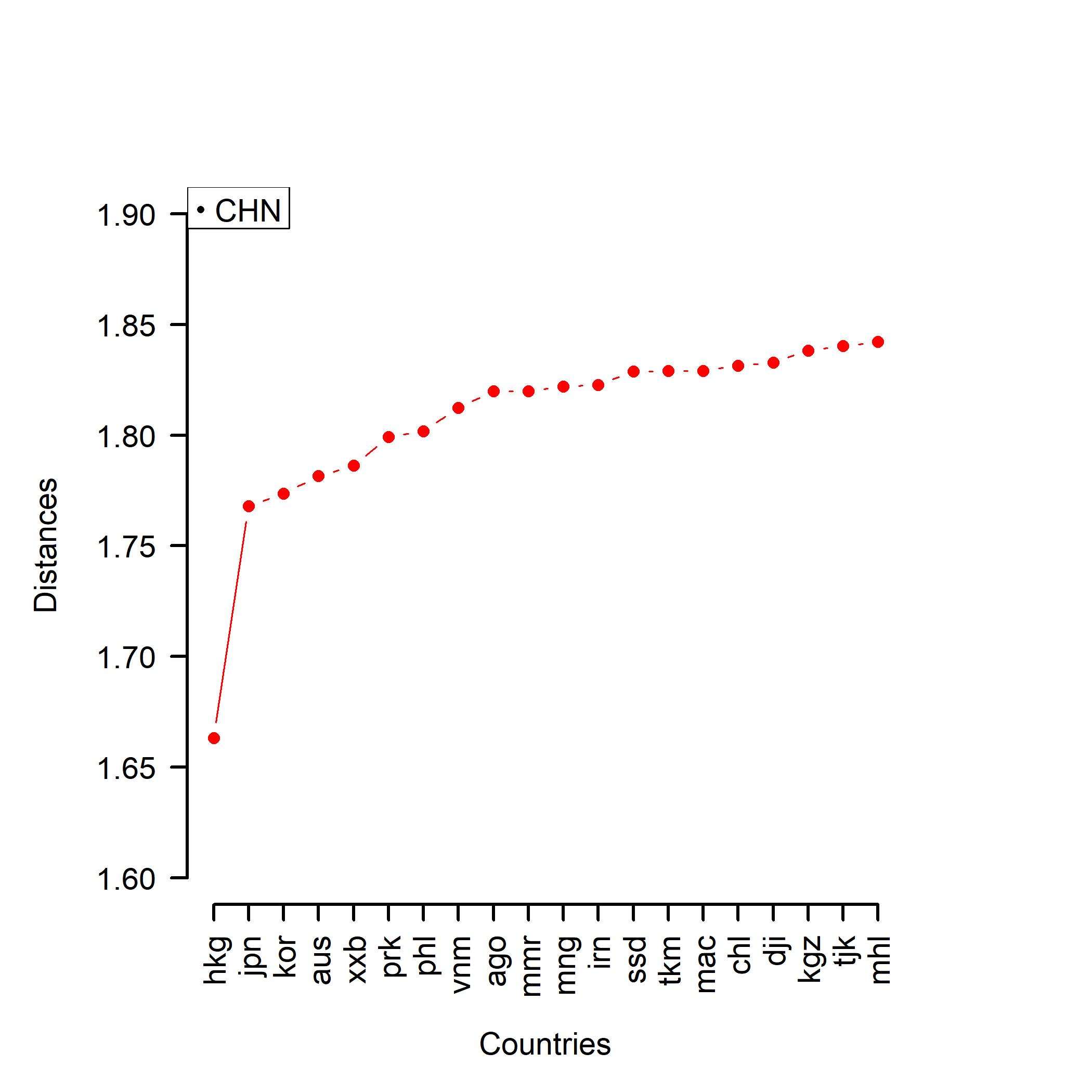}
			\includegraphics[width=.49\linewidth]{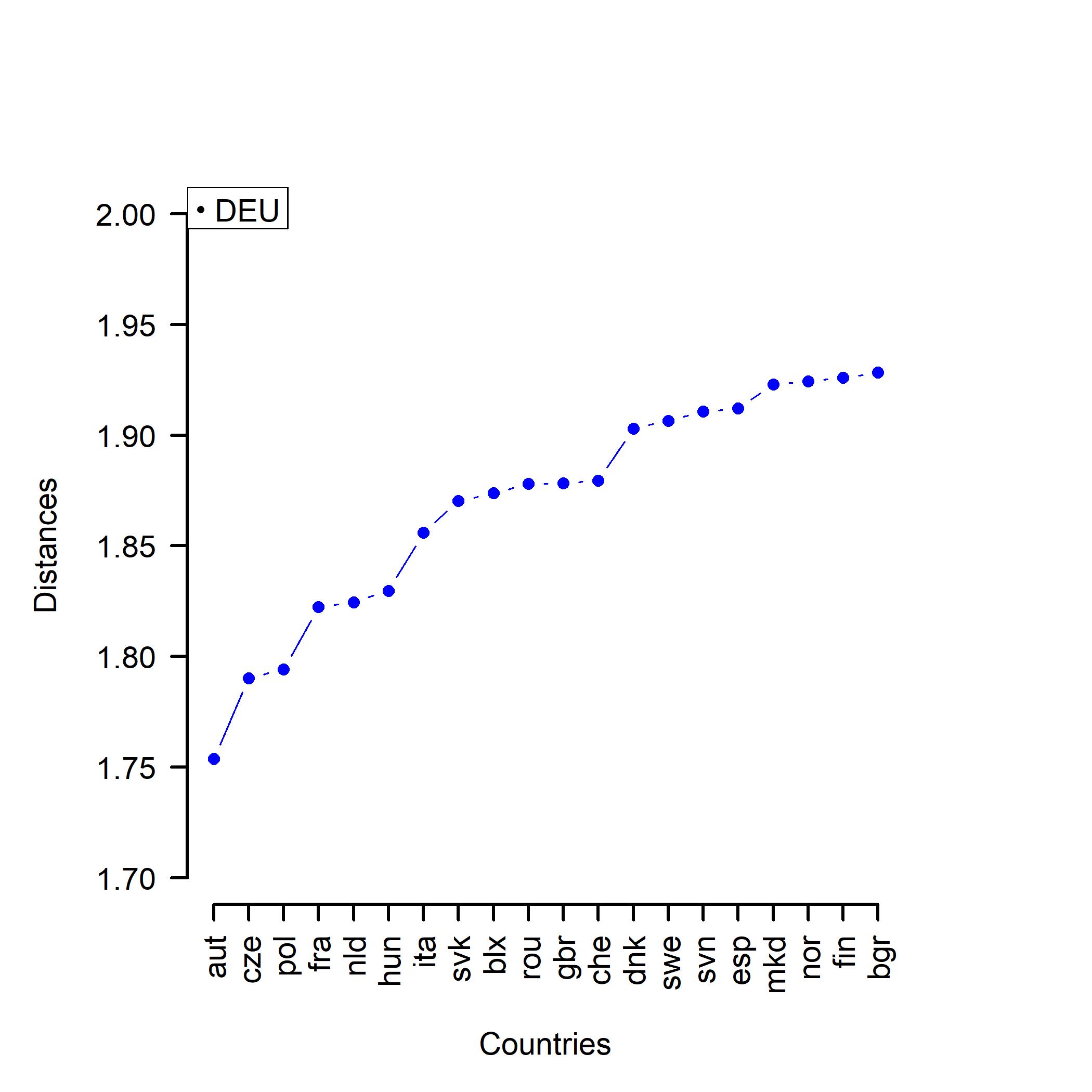}
			\caption{Top 20 nearest countries for  China (left) and Germany (right)}
			\label{fig:22}
		\end{figure}

Finally, from the point of view a single country, it is worth to look for the closest trade partners, that is the nearest nodes in terms of resistance distance. Figure \ref{fig:22} shows the distance profiles for the most central country of community 1 and 2, respectively. These plots can be interpreted as the list, in decreasing order, of countries that are most positively correlated with the selected centre, China or Germany. For instance, while in terms of communicability distance China is well-communicating with USA (third position in Figure \ref{fig:14}), USA does not belong to the top 20 most correlated countries with China. Rather, the left-hand side in figure \ref{fig:22} clearly shows a driving and synchronizing effect of the Chinese giant in the entire South-East Asia area. Similarly, figure \ref{fig:22} (right-hand side) confirms the role of Germany in the European Union and the strong correlation with Austria, Czech Republic and Poland.
				
	\hfill\eject
	
\subsection{Comparison with other approached applied to the same network}

It is worth briefly comparing our results with those obtained by other methodologies on the same network (see \cite{Barigozzi2011} and \cite{Piccardi2012}). In particular, in \cite{Piccardi2012}, several approaches are proposed to analyse the community structure of the WTN at different times. The authors showed that the recognition of mesoscale structures is increasingly difficult also because the world is becoming increasingly global over time. This makes even more compelling the search for a method that forces even slight deviations from a random structure to emerge. 
	Both directed and undirected networks have been tested, although no significant differences have been found.
	As in our case, results reported in \cite{Piccardi2012} show that geographical proximity still matters for international trade, jointly with trade agreements, common language or religion, and traditional partnerships. In particular, focusing on the application of a classical maximum modularity criterion, the authors find in 2008 (the most recent year of their analyses) three big communities containing $68$, $66$, and $47$ countries, with the largest cluster associated with Asia and Oceania. This is partially in line with our result in which a large relevant community including China, Oceania and North America is observed. On the other hand, by using either communicability or resistance distance, we found a higher level of granularity. Additionally, our approach provides a higher flexibility allowing to emphasize stronger connections when the threshold increases.

	The authors in \cite{Piccardi2012} also adopt a notion of distance among nodes based on random walks by row-normalizing the weight matrix. Modelling the WTN by stochastic matrix corresponds to moving from absolute to relative trade values. That distance between nodes is defined by complementing a similarity measure. A dendrogram is computed initially by defining groups containing single nodes and then by iteratively linking pairs of groups with minimal distance. This approach looks similar to ours being based on a varying threshold. They choose to maximize the so-called cophenetic correlation coefficient, which is defined as the linear correlation between the distances and the cophenetic distances, which are the heights of the link joining (directly or indirectly) nodes in the dendrogram. \\
	Some common evidences are noticeable also in this case. The United States and Canada form one of the strongest partnerships: their distance in the dendrogram stays constantly very small over time. France is strongly connected to some of its former colonies, as we also pointed out above, whereas Germany is close to other European countries. 
	Main differences are related to the behaviour of very small countries. While in our case, small countries are often classified as isolated nodes. In \cite{Piccardi2012}, very small countries are connected to much larger ones as an effect of the disassortativity observed in the WTN. These links tend to be small in absolute terms, given the small economic size of the countries, but they appear as relevant in relative terms, because the strong preference for a given partner.
	
The authors in \cite{Piccardi2012} also used stability and persistence to confirm their results. A random walker starting in a community is likely to remain for quite a long time within that community, before leaving it to enter another one. The analysis of the persistence probabilities induced in a network by a given partition has recently been proven to be an effective tool for testing the existence and significance of communities. Also in this case, we observe that communities with high persistence probability have common features with our results. Indeed, the top communities identified in \cite{Piccardi2012} considers the entire set of European countries, plus a number of minor non-European partners, that is in line with the top community selected by the communicability approach. Similarly, the second large community with a high persistence probability includes the entire North America and most of Central and South America, plus China, Australia, and many others. Although less granular, this community is fully comparable with community 2 detected by the communicability approach.
	
A quantitative correlation between the world partition in communities obtained by a modularity criterion and geographical distances has been investigated in \cite{Barigozzi2011}. The authors, both at an aggregate level and at a number of  commodity-specific levels, compare the two maximum modularity partitions of the input-output network and of the weighted network of the geographical closenesses. They find a high similarity between aggregate trade and geography-based communities, greater than, for instance, communities determined by regional trade agreements. They conclude that geographically-related factors explain the patterns of global trade more than political determinants. Although a positive correlation is present between monetary flows and geographical closenesses, we noticed that the geographical distances are less relevant when indirect relationships are also considered via either communicability or resistance distances\footnote{The rank correlation between these distances and the geographical distance between capital cities is lower than 0.15.}. As a consequence, the community structure we find appears more granular than the groups found in \cite{Barigozzi2011} and the composition cannot be explained only by geographical patterns. Other factors are involved as historical  relationships, trade agremeents and strategic economic alliances.
	
To conclude, although some common results with \cite{Barigozzi2011} and \cite{Piccardi2012} are observed, our methodology has the advantage of clearly highlighting even small differences and forcing the emergence of very strong ties between different countries through the use of a distance threshold. Furthermore the modularity $Q$ we applied turns out to be a simple and flexible tool, more homogeneous to the context of a network interpreted as a metric space.

		\section{Conclusions and further research}
		\label{sec:9}
		Community detection is a key topic in the analysis of complex systems, where discovering the inner structure plays a relevant role. 
		In particular, the centrality of countries and the relationships between them assume specific relevance in the World Trade Network, where economical and geopolitical phenomena affect over time the structure of the global network. In this framework, this work aimed at detecting different levels of clustered communities in the network on the basis of both communicability and resistance distances. 
		The proposed methodology allows to discover the hidden hierarchical structure of the network, as it presents a degree of flexibility highlighting very tight relationships by varying the threshold parameter, and revealing in this way the clusters of nodes that more easily communicate. Moreover, it performs well also for weighted and extremely dense network, as the case of the WTN. \\
		Features and properties of each community can be exploited in order to compare the characteristics of different clusters and to detect the most central countries inside the single community as well in the whole network. \\
		Numerical results depict the structure of the economic trade detecting main relevant communities. In particular, main community sees United States and China as main actors. Most flows are polarized around the exchange channel between China and USA and all their satellite countries. However, focusing on the correlation between trades, the procedure emphasizes the different role of these two countries. In particular, it is worth mentioning the emerging of China-Oceania community when deep links emerge. Furthermore, it is confirmed that Germany plays a key role in Europe and preferential channels of internal exchanges are observed in the European market. 
		 In line with \cite{Zhu2014}, emphasizing tight links, we obtain that although the strong trade relationships with USA and Germany, China became regionally attractive and restored the leadership of Asia-Oceania community. European community is highly centralized around founding  members of the European Economic Community with the central role of Germany. High income countries in Northern Europe are instead in a separate community with a less relevant role in the network.

\clearpage

\bibliographystyle{spmpsci}

\bibliography{References}

\clearpage

\begin{appendices}

\section*{Appendix A}
\label{AppendixA}

We report here thorough computations and proofs of formulae \ref{potential}, \ref{partition} and \ref{vibrational2} in section \ref{sec:4.2}. The expression \ref{potential} of the total potential energy $U$ can be handled in the following way: 

\begin{equation*}
\begin{split}
U&=
\frac{1}{4}{\cal K}\sum_{i, j}A_{i j}[z^2_i-2z_i z_j+z^2_j]\\
&=\frac{1}{4}{\cal K}\left[\sum_{i}z_{i}^2\sum_{j} A_{i j}- 2\sum_{i, j}A_{i j}z_i z_j + \sum_{j}z_{j}^2\sum_{i} A_{i j}\right]\\
&=\frac{1}{4}{\cal K}\left[\sum_{i}z_{i}^2k_{i}- 2\sum_{i, j}A_{i j}z_i z_j + \sum_{j}z_{j}^2k_{j}\right]=\frac{1}{2}{\cal K}\left[\sum_{i}z_{i}^2k_{i}- \sum_{i, j}A_{i j}z_i z_j\right]\\
&=\frac{1}{2}{\cal K}\left[\sum_{ij}z_{i}(\textbf{K-A})_{ij} z_j\right]=\frac{1}{2}{\cal K}\left[\sum_{ij}z_{i}L_{ij} z_j\right].
\end{split}
\end{equation*}


We compute now the expression of the partition function $\cal Z$ in formula \ref{partition}. Using the spectral decomposition of the Laplacian matrix $\textbf{L}=\textbf{M}\bf{\Lambda}\textbf{M}^T$, where $\bf{\Lambda}$ is the diagonal matrix of the eigenvalues and $\textbf{M}$ is the corresponding matrix of the eigenvectors, we have the following chain of equalities:

\begin{equation*}
\begin{split}
{\cal Z}
&=\int e^{-\frac{1}{2}\beta {\cal K}\sum_{ij}z_{i}{(\textbf{M}{\bf \Lambda} \textbf{M}^T)_{ij}} z_{j}}\prod_{k}dz_k=\int e^{-\frac{1}{2}\beta {\cal K}\textbf{z}^{T} (\textbf{M}{\bf \Lambda} \textbf{M}^T)\textbf{z}}\prod_{k}dz_k \\
&=\int e^{-\frac{1}{2}\beta {\cal K}({\textbf{M}^T}\textbf{z})^T {\bf \Lambda} (\textbf{M}^T \textbf{z})}\prod_{k}dz_k=\int e^{-\frac{1}{2}\beta {\cal K}\textbf{x}^T {\bf \Lambda} \textbf{x}}\prod_{k}dx_k\\
&=\int e^{-\frac{1}{2}\beta {\cal K} \sum_{k}\mu_{k}x_{k}^2}\prod_{k}dx_k\\
\end{split}
\end{equation*}

where we set $\textbf{x}=\textbf{M}^T \textbf{z}$ and $d\textbf{z}=|\det \textbf{M}|d \textbf{x}=d \textbf{x}$.\\
As usual in literature, we remove the contribution from $\mu_n=0$, providing the modified partition function we still call $\cal Z$, (see \cite{Estrada2012book} pag. 117); this yields:

\begin{equation*}
\begin{split}
{\cal Z}
&=\prod_{k=1}^{n-1}\int e^{-\frac{1}{2}\beta {\cal K}\mu_{k}x_{k}^2}dx_k=\prod_{k=1}^{n-1}\sqrt{\frac{2\pi}{\beta {\cal K}\mu_k}}
\end{split}
\end{equation*}

the last equality being valid because all integrals are Gaussian with $\mu_{k} > 0$, $k=1,...,n-1$.


Finally we compute $G^{v}_{ij}(\beta)$ (formula \ref{vibrational2}):

\begin{equation*}
\begin{split}
G^{v}_{ij}(\beta)
&=\frac{1}{\cal Z}\int z_i z_j e^{-\beta U} d\textbf{z}\\
&=\dfrac{1}{\cal Z}\int z_i z_j e^{-\frac{1}{2}\beta {\cal K} \sum_{ij}z_{i}L_{ij} z_j} \prod_{k}dz_k\\
&=\dfrac{1}{\cal Z}\int (\textbf{M}\textbf{x})_i (\textbf{M}\textbf{x})_j e^{-\frac{1}{2}\beta {\cal K} \sum_{k}\mu_{k}x_{k}^2} \prod_{k}dx_k\\
&=\dfrac{1}{\cal Z}\int \left(\sum_{k=1}^{n}\psi_{k}(i)x_k\right) \left(\sum_{k=1}^{n}\psi_{k}(j)x_k\right) \prod_{k=1}^{n} e^{-\frac{1}{2}\beta {\cal K}\mu_{k}x_{k}^2} dx_k\\
\end{split}
\end{equation*}


Notice that, computing the product of the two sums inside the integral above, all the integrals involving mixed terms are null, as the integrand is an odd function and the integral is extended to  $\mathbb R$ for each $x_k$. Then, only the squared terms remain inside the integral, so that: 

\begin{equation*}
\begin{split}
G^{v}_{ij}(\beta)
&=\dfrac{1}{\cal Z}\int \big(\psi_1(i)\psi_1(j)x_1^2+\dots  +\psi_n(i)\psi_n(j)x_n^2 \big) \prod_{k=1}^{n} e^{-\frac{1}{2}\beta {\cal K}\mu_{k}x_{k}^2} dx_k\\
&=\dfrac{1}{\cal Z} \psi_1(i)\psi_1(j) \int x_1^2 e^{-\frac{1}{2}\beta {\cal K}\mu_{1}x_{1}^2}dx_1 \cdot \int e^{-\frac{1}{2}\beta {\cal K}\mu_{2}x_{2}^2}dx_2 \cdot \ \dots \ \cdot  \int e^{-\frac{1}{2}\beta {\cal K}\mu_{n}x_{n}^2}dx_n+\dots +\\
&\ \ \ \ \dfrac{1}{\cal Z}\psi_n(i)\psi_n(j) \int e^{-\frac{1}{2}\beta {\cal K}\mu_{1}x_{1}^2}dx_1 \cdot \int e^{-\frac{1}{2}\beta {\cal K}\mu_{2}x_{2}^2}dx_2 \cdot \ \dots \ \cdot  \int x_n^2 e^{-\frac{1}{2}\beta {\cal K}\mu_{n}x_{n}^2}dx_n
\end{split}
\end{equation*}

We remove once again the contribution from $\mu_n=0$, then computing the integrals we have:

\begin{equation*}
\begin{split}
G^{v}_{ij}(\beta)
&=\dfrac{1}{\cal Z} \psi_1(i)\psi_1(j)\frac{\sqrt{2\pi}}{\sqrt{(\beta {\cal K}\mu_1)^3}}\cdot \frac{\sqrt{2\pi}}{\sqrt {\beta {\cal K}\mu_2}}\cdot \ \dots \ \cdot \frac{\sqrt{2\pi}}{\sqrt {\beta {\cal K}\mu_{n-1}}}+ \dots +\\
&\ \ \ \ \dfrac{1}{\cal Z} \psi_{n-1}(i)\psi_{n-1}(j) \cdot \frac{\sqrt{2\pi}}{\sqrt {\beta {\cal K}\mu_1}} \cdot \frac{\sqrt{2\pi}}{\sqrt {\beta {\cal K}\mu_2}}\cdot \ \dots \ \cdot \frac{\sqrt{2\pi}}{\sqrt{(\beta {\cal K}\mu_{n-1})^3}}\\
&=\frac{1}{\cal Z} \prod_{k=1}^{n-1}\sqrt{\frac{2\pi}{\beta {\cal K}\mu_k}} \left[ \frac{\psi_1(i)\psi_1(j) }{\beta {\cal K}\mu_1}+ \dots + \frac{\psi_{n-1}(i)\psi_{n-1}(j) }{\beta {\cal K}\mu_{n-1}}\right]=\sum_{k=1}^{n-1}\frac{\psi_k(i)\psi_k(j)}{\beta {\cal K}\mu_k}.
\end{split}
\end{equation*}



\section*{Appendix B}
\label{AppendixB}

The expression of the pseudo-inverse of the Laplacian $\textbf{L}^+=\left(  \textbf{L}+ \frac{1}{n}  \textbf{J} \right)^{-1} -\frac{1}{n} \textbf{J}$ allows an interesting interpretation of the resistance distance $\omega_{ij}$ in an economic, or financial, networked system.

Suppose that, to each node, a value of a given attribute is assigned through a state vector $\textbf{v}=[v_1, v_2, \dots, v_n]^T$ (such an attribute could be, for instance,  the GDP of a Country or the assets of a financial institution), and let $I_{ij}=v_i -v_j$ be the flow of such an attribute from node $i$ to node $j$.
We denote by $I_i$ the total outgoing flow from the node $i$ to its adjacent nodes, i.e. $I_i=\sum_{j=1}^{n}a_{ij}(v_i -v_j)$.\\ 
In matrix form, the total outgoing flow of the nodes attribute is then $$\bf I=(K-A)v=Lv.$$ 
The Laplacian matrix transforms nodes attributes $v_i$, $i=1,...,n$ into outgoing flows from nodes $I_i$, under the assumption that a flow $I_{ij}$ along a given edge is equal to the gradient $\Delta v_{ij}=v_{i}-v_{j}$. 
This assumption is equivalent to choose an effective resistance equal to $1$ along all edges. Of course, we may have both outgoing and ingoing currents according to the sign of $\Delta v_{ij}$: positive for outgoing flows from $i$ and negative for ingoing flows into $i$.\\
A similar meaning can be given to $\left( \textbf{L}+\frac{1}{n} \textbf{J} \right)\textbf{v}$. Indeed, $$\left( \textbf{L}+\frac{1}{n} \textbf{J} \right)\textbf{v}=\textbf{Lv}+\frac{1}{n} \textbf{Jv}=\textbf{I}+\overline{v}\textbf{u},$$ where  $\overline{v}=\frac{1}{n}\sum_{k=1}^{n}v_{k}$, that is, the operator $\textbf{L}+\frac{1}{n} \textbf{J}$ adds to the flows a constant term given by the mean value of all the attributes of the nodes. 
Then, the matrix $\left( \textbf{L}+\frac{1}{n} \textbf{J} \right)$ transforms nodes attributes $\textbf{v}$ into total outgoing flows $\textbf{I}$ 
in the network, up to an additive constant.\\
In a similar way, the inverse $\left( \textbf{L}+ \frac{1}{n} \textbf{J} \right)^{-1}$ acts on a current vector $\textbf{I}$ and produces a state vector $\textbf{v}$, which can be interpreted as the cause of such currents in the network. Specifically

\begin{equation*}
\textbf{v}={\textbf{L}^+}\textbf{I}= \left[\left( \textbf{L}+ \frac{1}{n} \textbf{J} \right)^{-1} -\frac{1}{n}\textbf{J} \right] \textbf{I}=\left( \textbf{L}+ \frac{1}{n} \textbf{J} \right)^{-1}\textbf{I}-\overline{\rm I}\textbf{u}
\end{equation*}


where, once again, the term $\frac{1}{n}\textbf{JI}=\overline{\rm I}\textbf{u}$ is the average value of the outgoing currents coming from every node.

Suppose now that in the system there are an outgoing flow equal to $1$ from a node (node $1$, for instance), an ingoing flow equal to $-1$ into another node (for instance, node $2$), whereas for all the other nodes the flow is zero. 
This is equivalent to a current vector equal to $\textbf{I}=[1, -1, 0, \dots, 0]^T=\textbf{e}_1-\textbf{e}_2$. Loosely speaking, a unit information is coming out from node $1$ and goes entirely into node $2$. To produce these flows, we have to start from an initial attributes vector on nodes given by

\begin{equation*}
\textbf{v}={\textbf{L}^+}(\textbf{e}_1-\textbf{e}_2)= \left[\left( \textbf{L}+ \frac{1}{n} \textbf{J} \right)^{-1} -\frac{1}{n}\textbf{J} \right] \left( \begin{array}{c} 
1   \\
-1  \\
0   \\
\vdots  \\
0 \\
\end{array} \right)= \left( \textbf{L}+ \frac{1}{n} \textbf{J} \right)^{-1}\left( \begin{array}{c} 
1   \\
-1  \\
0   \\
\vdots  \\
0 \\
\end{array} \right)
\end{equation*}

where the last equality holds because $\textbf{J}(\textbf{e}_1-\textbf{e}_2)=\textbf{0}$, that is $\overline{\rm I}=0$. 
Thus, the resistance distance between nodes $1$ and $2$ is  given by $$\omega_{12}=(\textbf{e}_1-\textbf{e}_2)^T\left( \textbf{L}+ \frac{1}{n} \textbf{J} \right)^{-1} (\textbf{e}_1-\textbf{e}_2)=v_1-v_2=\Delta v_{12}.$$

If $\Delta v_{12}$ is small, a small gradient is enough to transmit such a unit flow from node $1$ to node $2$; whereas, if $v_1-v_2$ is big, a high gradient is needed in order to produce the same unit flow. More in general, let's imagine that in the node $i$ the value ${v_i}$ is positive. Then the fact that another attribute $v_j$ with $j\neq i$ is positive means that node $i$ and node $j$ are strongly correlated since it is enough a low attribute difference to subtract from node $i$ a unit flow. This means that these two nodes communicate a lot. Whereas, if for another node $k$ with $k \neq i$, the corresponding component $v_k$ is negative this implies that node $i$ and node $k$ are strongly anti-correlated since, in order to produce a unit flow from node $i$, node $k$ has to be at a negative attribute, i.e. the attribute difference between $i$ and $k$ must be high. This means that the two nodes don't communicate well. The signs of the components of the vector $\textbf{v}$ indicate nodes that are positively or negatively correlated with node $i$ according to the fact these components have the same sign as $v_i$ or not. Let us observe that, in general, $\textbf{v}={\textbf{L}^+} \textbf{I}={\textbf{L}^+} (\textbf{e}_i -\textbf{e}_j)=L^+_i- L^+_j$ with $L^+_i$ $i$-th column of the matrix $L^+$. That is, if we want to decrease by $1$ the attribute of node $i$ and increase by $1$ the attribute of node $j$, we have to take an initial distribution of attributes on nodes equal to the difference between $i$-th column of $\textbf{L}^+$ and $j$-th column of $\textbf{L}^+$, and these columns are also the values of vibrational communicability $\textbf{G}^{v}$ between nodes, as defined in the text.

\end{appendices}

\end{document}